\documentclass[a4paper, 12pt]{amsart}
\usepackage{pdfsync}
\usepackage[english]{babel}
\usepackage{enumitem}
\setlist[itemize,1]{label=\textbullet}
\usepackage[T1]{fontenc}
\usepackage{fancyvrb} 

\usepackage{amsmath}
\usepackage{amssymb}
\usepackage{amscd} 
\usepackage{xcolor}

\usepackage{listings}

\lstdefinestyle{mypython}{
  language=Python,
  basicstyle=\ttfamily,
  keywordstyle=\color{blue},
  commentstyle=\color{gray},
  stringstyle=\color{orange},
  showstringspaces=false
}

\lstnewenvironment{code}
  {\lstset{style=mypython}}
  {}

\usepackage{amsthm}
\usepackage{mathrsfs}
\DeclareMathAlphabet{\mathpzc}{OT1}{pzc}{m}{it}

\newtheorem{theorem}{Theorem}
\newtheorem{lemma}[theorem]{Lemma}
\newtheorem{corollary}[theorem]{Corollary}

\theoremstyle{definition}
\newtheorem{definition}[theorem]{Definition}
\newtheorem{remark}[theorem]{Remark}
\newtheorem{example}[theorem]{Example}

\usepackage[colorlinks=true, linkcolor=gray, citecolor=red, urlcolor=magenta]{hyperref}

\title[Noisefree homomorphic encryption]{Introducing GRAFHEN: \underline{Gr}oup-b\underline{a}sed \underline{F}ully \underline{H}omomorphic \underline{E}ncryption without \underline{N}oise}

\author[Guillot]{Pierre Guillot ${}^1$}
\thanks{${}^1$ \texttt{pierre.guillot@raveltech.io}}
\thanks{${}^2$ \texttt{auguste.hoangduc@raveltech.io}}
\thanks{${}^3$ \texttt{michel.koskas@raveltech.io}}
\thanks{${}^4$ \texttt{florian.mehats@raveltech.io}}

\author[Hoang Duc]{Auguste Hoang Duc ${}^2$}

\author[Koskas]{Michel Koskas ${}^3$}

\author[Méhats]{Florian Méhats ${}^4$}

\date{}

\newcommand{\pierre}[1]{}
\newcommand{\florian}[1]{}

\newcommand{\Dec}{\mathrm{Dec}}

\newcommand{\Gc}{\mathscr{G}}

\newcommand{\Zca}{\mathcal{Z}}

\newcommand{\F}{\mathbb{F}} \newcommand{\K}{\mathbb{K}}
\newcommand{\Z}{\mathbb{Z}}

\newcommand{\SL}{\operatorname{SL}} \newcommand{\PSL}{\operatorname{PSL}}
\newcommand{\add}{\mathpzc{add}} \newcommand{\mul}{\mathpzc{mul}}
\newcommand{\enc}{\operatorname{Enc}} \newcommand{\dec}{\operatorname{Dec}}

\renewcommand{\phi}{\varphi}

\renewcommand{\bar}{\overline}

\newcommand{\smb}[1]{\langle #1 \rangle}
\newcommand{\psm}{\hat}  

\newcommand{\finex}{\hfill $\square$}
\renewcommand{\epsilon}{\varepsilon}

\newcommand{\rws}{rewriting system}
\newcommand{\rwss}{rewriting systems}
\newcommand{\lrs}{\longleftrightarrow^*} 

\newcommand{\aut}{\operatorname{Aut}}

\newcommand{\runningexample}[1]{%
\smallskip\noindent\hspace{.15\textwidth} \begin{minipage}{.8\textwidth}   
\footnotesize #1
\end{minipage}
}

\usepackage[capitalise, nameinlink]{cleveref}
\usepackage{graphicx}

\usepackage{xspace}

\begin{document}

\maketitle

\vspace*{-2mm}
\begin{center}
Ravel Technologies, Paris
\end{center}
\bigskip

\begin{abstract}
We present GRAFHEN, a new cryptographic scheme which offers Fully Homomorphic
Encryption without the need for bootstrapping (or in other words, without
noise). Building on the work of Nuida and others, we achieve this using
encodings in groups.

The groups are represented on a machine using {\em rewriting systems}. In this
way the {\em subgroup membership problem}, which an attacker would have to solve
in order to break the scheme, becomes maximally hard, while performance is
preserved. In fact we include a simple benchmark demonstrating that our
implementation runs several orders of magnitude faster than existing standards.

We review many possible attacks against our protocol and explain how to protect the scheme in each case.

An appendix describes an expert's attempts at cracking the system.
\end{abstract}

\section{Introduction}
\noindent Fully Homomorphic Encryption (FHE) has emerged as a cornerstone of
modern cryptography, enabling computation on encrypted data without requiring
access to the plaintext. Since Gentry’s seminal 2009 construction \cite{gentry},
FHE has promised powerful applications in privacy-preserving computation, secure
outsourcing, and confidential machine learning. However, despite its theoretical
potential, practical deployment has remained limited by the high computational
cost associated with managing noise in ciphertexts during evaluation.

In most FHE schemes—particularly those based on lattice problems such as
Learning With Errors (LWE) \cite{regev} and Ring-LWE \cite{lpr}—the encryption
process deliberately injects noise into ciphertexts to ensure semantic security.
As homomorphic operations are performed, this noise grows, and when it exceeds a
certain threshold, the ciphertext becomes undecryptable. Thus, the ability to
perform arbitrary computations is tightly constrained by the need to control
noise growth.

To address this, Gentry introduced the concept of bootstrapping—a technique
whereby a ciphertext is homomorphically refreshed by evaluating its own
decryption circuit, thereby reducing the accumulated noise. This method enables
unbounded-depth computation but at a significant computational cost, as the
bootstrapping step is typically the most expensive part of the pipeline.

In response, two broad strategies have emerged to mitigate the limitations of
bootstrapping. The first consists in avoiding bootstrapping altogether by
developing leveled FHE schemes, such as BGV \cite{bgv}, BFV \cite{bfv1,bfv2},
and CKKS \cite{ckks1,ckks2}. These constructions allow the evaluation of
circuits of limited depth without noise refreshing, relying instead on
techniques like modulus switching and key switching to control noise growth.
However, their expressiveness remains fundamentally bounded by the initial noise
budget and the noise growth rates inherent to the underlying operations.

The second strategy embraces bootstrapping but focuses on making it faster.
Notable examples include FHEW \cite{dm}, TFHE \cite{cggi1,cggi2}, which
introduced significant algorithmic and implementation-level optimizations to
drastically reduce the cost of bootstrapping. These schemes make it feasible to
perform a bootstrapping operation after every gate, opening the door to
low-latency homomorphic evaluation of Boolean circuits. While this represents a
substantial practical improvement, the need for frequent bootstrapping still
imposes a structural complexity and performance overhead that can be prohibitive
in large-scale applications. Recently, this line of work has been extended to
operate on large integers, rather than just bits. In particular,
\cite{rhe1,rhe2} significantly broadens the scope of bit-level
bootstrapping-based approaches, bringing them closer to practical applications
in encrypted numerical computing.

In summary, despite more than a decade of progress, noise remains a central
challenge in homomorphic encryption. Existing schemes must either limit the
depth of supported computations or incur the cost of frequent bootstrapping.
Bridging this gap—enabling arbitrary-depth computation without the burden of
noise management—remains a key objective in the ongoing development of
efficient, fully homomorphic encryption.

As Gentry emphasized in his 2022 ICM survey \cite{gentryICM}, the notion of a
truly noise-free FHE—where ciphertexts remain decryptable after arbitrary
computations without any need for noise control—remains largely unfulfilled. He
notes that if such a scheme were to exist, it would likely look very different
from today’s lattice-based frameworks. Some speculative directions include
building FHE from multilinear maps \cite{boneh}, which generalizes bilinear
pairings and would allow for direct composition of encrypted functions without
error accumulation. Another avenue involves using non-solvable group structures
\cite{Nuida2020}, where homomorphic operations might be encoded through group
actions that naturally avoid noise growth. However, both approaches remain
largely theoretical and face major barriers, including unresolved issues of
security, functionality, or concrete efficiency. 

In this paper, we build on the work of Nuida in \cite{Nuida2020}, which in turn
relied on pioneering work by Ostrovsky {\em et al.} \cite{Ostrovsky2007}. A fundamental
observation, which we describe at length in \S\ref{section-defs}, is that fully
homomorphic encryption is possible if we can homomorphically encrypt the
elements of a carefully chosen group $E$. Thus the encrypted messages are
elements of another group $G$, and the decryption map is a group homomorphism to
$E$. The challenge is to choose the group $G$ (together with a machine
implementation) so that the resulting scheme is secure. The fundamental problem
which needs to become hard for an attacker and remain easy for the user is the
{\em subgroup membership problem}, that is, the problem of deciding whether an
element $g\in G$ belongs to a given subgroup $Z$ or not. This depends
drastically on the representation of $G$, and is trivial for permutation groups
for example.

Finding the right type of group  was an open question at the end of
\cite{Nuida2020}, and we propose a solution, which is at the heart of our
cryptosystem named GRAFHEN, for GRoup-bAsed Fully Homomorphic Encryption without
Noise. The basic idea is to work with {\em rewriting systems} (the theory of
which is summarized in \S\ref{sec-rws}). Given a choice of generators $x_1,
\ldots, x_d$ for a group $G$, a rewriting system is a collection of rules of the
form $u \longrightarrow v$ where $u$ and $v$ are words on the alphabet $\{ x_1,
\ldots, x_d \}$, which hold true in $G$, and allow us to rewrite any word and
obtain a ``simpler'' one representing the same element of the group.
Traditionally this is used to obtain canonical forms for the elements of the
group; for us however, the rules are chiefly a way of obtaining words of bounded
length (the rules $u \longrightarrow v$ being chosen with $v$ shorter than $u$).

Concretely, the generators $x_1, \ldots, x_d$ are to remain secret, and
constitute the private key. The rules are public. Users work with words and
rules. They have only access to the group $\tilde G$ defined ``by generators and
relations'' by the alphabet and the rules, and the subgroup membership problem
is known to be hard, in fact undecidable in general, for such groups. For more
about this, see the beginning of \S\ref{sec-attacks}.

We shall review the main attacks on our scheme, and argue that with sensible
choices of parameters, it becomes fully secure. We also include an appendix,
written by James Mitchell who is a specialist in computational group theory (and
aguably the best expert in the world on the specific task at hand), in which he
describes his attempts at breaking GRAFHEN: while this can be done for tiny
values of the security parameters, it becomes rapidly impossible.  Thus we offer
a secure protocol for fully homomorphic encryption without noise (or
bootstrapping).

Here we want to address a point. GRAFHEN messages are words, which need to be
shortened every now and then using the rules. We would like to insist on the
differences between this rewriting and the usual bootstrapping. First, on a
theoretical level, a major difference is that the bootstrapping operation to
refresh the ciphers {\em must be performed} after a certain number of
homomorphic operations were made {\em even on a idealized computer with infinite
memory} -- otherwise the messages are lost. By contrast, we only need to shorten
the messages when we feel the pressure on the memory is too great -- and in
situations where memory usage is not critical, but speed is, we may decide not
to perform the reductions for a while (and perhaps wait for the CPU to be idle
to do this). In fact, the operation is one of {\em lossless data compression}.

Another difference is one of performance. Rewriting words does not involve
arithmetic operations, only some string matchings and replacements which can be
performed efficently (using automata for example); a bootstrapping may involve
millions of multiplications between floating point numbers\pierre{une réf pour
ça?}. We include a benchmark in \S\ref{sec-parameters} which shows that GRAFHEN is several orders of magnitude faster than OpenFHE, for example.

\bigskip\noindent {\bfseries Acknowledgements.} We are grateful to Florent
Hivert and James Mitchell for many discussions on questions of security. James
Mitchell has also written Appendix \ref{appendix-james}, in which he reports on
his attempts to break the scheme we propose here. We are also indebted to Jean-Baptise Rouquier for his careful proofreading and many suggestions.

In addition, we extend our thanks to Koji Nuida for his encouraging words about an early version of the manuscript, and for spotting a typo. Finally, we have benefited from the advice of David Naccache, and we would like to express our gratitude.

\bigskip\noindent {\bfseries Some conventions.} When working with permutations,
we rely on some convenient conventions. We would like to see the symmetric group
$S_n$ as a subgroup of $S_{n+1}$. In order to achieve this, and following the
custom of certain computer algebra packages such as
GAP\footnote{\url{https://www.gap-system.org}}, we define a {\em permutation} to
be a bijection of $\Z$ which is the identity outside of a finite set, and $S_n$
is the group of permutations which are the identity outside of $\{ 1, \ldots, n
\}$. Thus $S_n\subset S_{n+1}$ as promised. An element such as $(15)$ (which can
be also written $(1,5)$ for clarity) belongs to $S_n$ for all $n\ge 5$. We note
that permutations, such as we have defined them, are best written in cycle
notation.

We also want to identify $S_n \times S_m$ as a subgroup of $S_{n+m}$. This we do
silently. That is, we may informally write $S_n\times S_m$ to mean the group of
all products $ab$ with $a\in S_n$ and $b$ a permutation which is the identity
outside of $\{ n+1, \ldots, n+m \}$. Of course this is only used in situations where
no confusion can arise.

We also point out, but this will only be of consequence (perhaps) for the
diligent reader who tries to reproduce all of our computations, that when $a$
and $b$ are permutations, we write $ab$ for $b\circ a$ -- that is, first apply
$a$, then apply $b$. This is again in order to follow GAP.

\tableofcontents

\section{Homomorphic encodings in groups} \label{section-defs}

In this section we describe the strategy laid out in \cite{Nuida2020}, itself
building on \cite{Ostrovsky2007}. We explain how the construction of a scheme
capable of encrypting a group element and to perform the group operation
homomorphically leads to a fully homomorphic encryption scheme. For this one
needs to pick what we call a {\em homomorphic encoding}, and we explore their
basic theory.

The notation introduced in this section remains in place throughout the paper.

\subsection{Goals \& strategy -- from two operations to one}

Let  $\K$ be a commutative ring. Many of our examples in the paper will deal
with $\K = \F_2 = \{ 0,1 \}$, the field with two elements, but some basic
principles hold more generally.

\begin{definition}
  A {\em homomorphic encryption of $\K$} is a map $\dec \colon C \longrightarrow
  \K$, where $C$ is a set, together with two other maps 
  \[ \smb\add, \smb\mul \colon C
  \times C \longrightarrow C \]
  chosen so that the following diagrams are
  commutative:
  \[ \begin{CD} C \times C @>\smb\add>> C \\
  @V{\dec\times\dec}VV            @VV{\dec}V \\
  \K \times \K @>+>> \K \end{CD} \qquad \qquad \qquad \begin{CD} C \times C
  @>\smb\mul>> C \\
      @V{\dec\times\dec}VV            @VV{\dec}V \\
      \K \times \K @>\cdot>> \K \end{CD} \]
 In other words, we require $\dec(\smb\add(x,y))= \dec(x) + \dec(y)$ and
 $\dec(\smb\mul(x,y)) = \dec(x) \cdot \dec(y)$ for all $x, y \in C$.  
\end{definition}

The set $C$ (the elements of which are called ciphertexts or ciphers) and the
maps $\smb\add$ and $\smb\mul$ are public, whereas the ``decryption map''
$\dec$ is confidential. Questions of security will be discussed later, but the
reader should keep in mind that we assume the presence of an ``attacker'' having
access to many elements $x \in C$ as well as the corresponding value $\dec(x)$,
and that they must not be able to predict the value of $\dec(y)$ for any new
element $y \in C$. In other words, we expect a Chosen Plaintext Attack (CPA),
and the goal is to achieve CPA security.

\begin{remark}
When $\K= \F_2$, we are talking about the encoding of a bit, in such a way that
all possible operations on one or two bits can be performed on the ciphers. (For
example the operation on one bit which exchanges 0 and 1, or in others words the
NOT operation, is $x \mapsto 1+x$, and can be performed on ciphers as $c \mapsto
\smb\add(u,c)$ where $u$ is a cipher of $1$.) However, it is well-known that any
function $\F_2^n \longrightarrow \F_2$, for any $n \ge 1$, can be realized by
composing binary and unary operations: algebraically, this is because any such
map can be written as a polynomial, and in computer science this fact is better
known as the universality of the NAND gate. Thus when we speak of a homomorphic
encryption of $\F_2$, we imply a fully homomorphic encryption scheme. However in
practice, if one has an efficient homomorphic encryption of $\F_p$ for some
prime $p$, it may be largely preferable to use it rather than write all numbers
in binary and so on.
\end{remark}

There are classical examples of encryption protocols which are compatible with a
single operation (such as RSA for example). In order to encrypt two operations,
in the above sense, the first step is to encode $\K$ in a group, that is to pick
an injective map $\enc \colon \K \longrightarrow E$ where $E$ is a group. There
is no loss of generality in assuming that $\enc(0)$ is the identity element $e$
of $E$. (In other contexts, the identity element of a group is often denoted by
1, but here there is a risk of confusion with the element $1 \in \K$.)

We need some vocabulary to describe the properties which $\enc$ must satisfy.

\begin{definition}
Let $G$ be a group and let $f \colon G^n \longrightarrow G$ be a function, where
$n$ is an integer. We say that $f$ is {\em polynomial} when it is constructed
entirely from the group law on $G$; in other words, when it is of the form
\[ f(x_1, \ldots, x_n) = a_0 x_{i_0} a_1 x_{i_1}a_2 \cdots  a_k x_{i_k} a_{k+1} \]
for a certain integer $k$, where the $a_0, \ldots, a_{k+1}$ are elements of $G$
(and the indices $i_j$ need not be distinct).
\end{definition}

\begin{remark}
In this definition we do not allow the use of the inverse operation in the group
$G$ (so that the definition could be given in a monoid). However, little change
would be brought by allowing inverses: in all practical examples, the group $G$
is finite, and for $x\in G$ of finite order $s$, one has $x^{-1} = x^{s-1}$.
Another remark is that constant functions are polynomial (this is the formula
above for $k=-1$, or you may take it as a convention if you prefer).
\end{remark}

\begin{definition} \label{def-hom-encoding}
  A {\em homomorphic encoding of $\K$} in the group $E$ is an (injective)
  encoding $\enc \colon \K \longrightarrow E$ together with two polynomial maps
  $\add, \mul \colon E^2 \longrightarrow E$ such that the restriction of
  $\add$ (resp.\ $\mul$) to $\enc(\K)$ corresponds to the sum (resp.\
  product) of $\K$. In other words, we require $\add(\enc(k), \enc(\ell)) =
  \enc(k+\ell)$ and $\mul(\enc(k), \enc(\ell)) = \enc(k\cdot\ell)$, for all
  $k, \ell \in \K$.
\end{definition}

We will provide many examples in the next subsection. Let us indicate at once,
however, that we are going to show that $E= \SL_3(\K)$ can be endowed with a
homomorphic encoding of $\K$, for any ring $\K$.

The following construction is central. Assume that $E$ is equipped with a
homomorphic encoding of $\K$, and that $L$ is any group on which we have
defined a surjective morphism
\[ \pi \colon L \longrightarrow E \, . \]
(The letter L is for ``lift''.) One can then construct a homomorphic encryption
of $\K$ from this, by putting $C = \pi^{-1}(\enc(\K))$ and $\dec = \enc^{-1}
\circ \pi$, and lifting $\add$ and $\mul$ to maps $\smb\add, \smb\mul \colon L^2
\longrightarrow L$ in the natural way. In more detail, recall that $\add$ is
assumed polynomial, so that we may write for $x_1, x_2 \in E$:
\[ \add(x_1, x_2) =  a_0 x_{i_0}  a_1 x_{i_1} \cdots  a_k
x_{i_k}  a_{k+1} \, , \]
with $ a_i \in E$. Thus it suffices to choose $\psm  a_i \in L$ such that
$\pi(\psm a_i) =  a_i$ and to put for $x_1, x_2 \in C$:
\[ \smb\add (x_1, x_2) = \psm  a_0 x_{i_0}\psm a_1 x_{i_1} \cdots\psm a_k x_{i_k}\psm a_{k+1} \, . \]
We then have $\pi(\smb\add(x_1, x_2)) = \add(\pi(x_1), \pi(x_2))$. Similar
considerations apply to $\mul$. It is clear that $\smb\add$ and $\smb\mul$ map $C^2$
into $C$, and that we are in the presence of a homomorphic encryption of $\K$. 

This construction reduces the problem to the choice of $L$ and $\pi$, that is,
we are left with the task of encrypting the elements of $E$ in a way which is
compatible with the group law.

\begin{remark}[vocabulary]
  It is convenient to slightly abuse the language, and to call $E$ the {\em
  group of encodings} or the {\em group of plaintext messages} (thus giving each
  element of $E$ the status of ``message'', even when it is not in $\enc(\K)$).
  The elements of $C$ are called {\em ciphers}. In most examples, $C$ is
  actually a subgroup of $L$, and we call it accordingly the {\em group of
  ciphers} in this case. When the cipher $x \in C$ verifies $\pi(x) = \enc(k)$
  for some $k \in \K$, or in other words $\dec(x) = k$, we call $x$ a {\em
  cipher of $k$}.

  The ciphers of $0$ are precisely the elements of the kernel of $\pi$, which
  will always be denoted by $Z$.
\end{remark}

\begin{remark}
  It is not truly necessary to assume that $L$ is a group, in the considerations
  above, nor even a monoid: any set equipped with an internal composition law --
  what is called a {\em magma} -- would suffice. We explore this possibility
  further when studying rewriting systems that are not confluent, which give
  rise to a composition law for which associativity is not guaranteed. One could
  similarly relax the assumptions on $E$, but no practical example has appeared
  so far.
\end{remark}

\subsection{Homomorphic encodings -- hands-on approach}

Here we give a few examples of encodings. We begin by mentioning two results which were already in the literature.

\begin{example}
  In \cite{Ostrovsky2007}, one can find formulae for encoding $\F_2$ into $E =
  A_5$, along with the corresponding functions $\add$ and $\mul$. In
  \cite{Nuida2020}, the reader will find alternative expressions for the same
  $E$, as well as a variant using $\F_3$. \finex
\end{example}

In both cases, the number of group operations needed for $\add$ and
$\mul$ is rather large. We offer the following family of examples, which
works for any commutative ring $\K$ and uses very few operations.

\begin{example} \label{ex-psl3p}
We encode the commutative ring $\K$ in $E=\SL_3(\K)$ as follows:
$$
\enc : \K \longrightarrow \SL_3(\K), \qquad  x \longmapsto \begin{bmatrix}
1 & 0 & x \\
0 & 1 & 0 \\
0 & 0 & 1 \\
\end{bmatrix} \, .
$$
Let us show how this encoding can be made homomorphic. We see easily that
$\enc(k) \cdot \enc(\ell) =  \enc(k + \ell)$ for all $k, \ell \in \K$, or in
other words, we can define $\add$ by $\add(x_1, x_2) = x_1 \cdot x_2$.

We describe the multiplication next. We define the matrices:
\[ 
h := \begin{bmatrix} 0&1&0 \\ 1&0&0 \\ 0&0&-1 \end{bmatrix} \, , \qquad 
g := \begin{bmatrix} -1&0&0 \\ 0&0&1 \\ 0&1&0 \end{bmatrix} \, .
\]
Then we check that for all $k, \ell \in\K$:
$$
g \cdot \enc(k) \cdot g^{-1} = \begin{bmatrix}
1 & -k & 0 \\
0 & 1 & 0 \\
0 & 0 & 1 \\
\end{bmatrix} \, ,
$$

$$
h \cdot \enc(\ell) \cdot h^{-1} = \begin{bmatrix}
1 & 0 & 0 \\
0 & 1 & -\ell \\
0 & 0 & 1 \\
\end{bmatrix} \, .
$$
For any two elements $U,V$ in a group, we write
$$
[U,V] = U \cdot V \cdot U^{-1} \cdot V^{-1}
$$
for their commutator. One verifies readily that
\[ 
[g \cdot \enc(k) \cdot g^{-1}, h \cdot \enc(\ell) \cdot h^{-1}] =  \enc(k \cdot \ell) \, ,
\]
for $k, \ell \in \K$. In other words, we may put $\mul(x_1, x_2)=
[gx_1g^{-1}, h x_2 h^{-1}]$, and $\mul$ is polynomial.

We conclude that the maps $\enc$, $\add$ and $\mul$ together form a
homomorphic encoding of $\K$. \finex
\end{example}

This example comes in simple variants. Thus, one may observe that $\SL_3(\K)$
can be replaced by $\PSL_3(\K)$; more generally, whenever there is a normal
subgroup $N$ of $E$, and the composition
\[ \begin{CD} \K @>\enc>> E @>>> E/N \end{CD} \] is injective, then one may
choose to work with $E/N$ in place of $E$. The fact that $\add$ and
$\mul$ can “induce” maps $(E/N)^2 \longrightarrow E/N$ is obvious if one
keeps in mind that these functions are polynomial.

Another variant, based on another general fact: if  $E$ is a subgroup of $E'$,
and if $E$ is endowed with a homomorphic encoding of $\K$, then so is $E'$.
Again, we extend $\add$ and $\mul$ to $E'$ by exploiting their
polynomial nature. For example, as the symmetric group $S_7$ contains a subgroup
isomorphic to $\PSL_3(\F_2)$, we deduce that $S_7$ can be used for the
homomorphic encoding of $\F_2$.

With a little more work, we can even obtain:

\begin{example} \label{ex-S6}
For practical reasons, it is useful to work with permutation groups of the
smallest possible degree. We can modify ``by hand'' the example of $S_7$ just
given and operate within $S_6$. The end result is this. Define $\enc(0)= Id$
(the identity permutation) and $\enc(1)=(15)(34)$. We put $\add(x,y) = xy$ and
$\mul(x,y) =  a_1x a_1 a_2y a_2 a_1x a_1 a_2y a_2$ where $ a_1=(12)(56)$ and
$a_2=(35)$. The reader can check that this is a homomorphic encoding of $\F_2$
in $S_6$ -- there are, after all, very few verifications to make. \finex
\end{example}

\begin{remark} \label{rmk-five-mul}
In this last example, assuming $a_1a_2$ is precomputed, it takes 5 group
operations to obtain $\mul(x,y) = [(((a_1x) a_1a_2)y)a_2]^2$. One can show that
this optimal, at least in the following sense: a homomorphic encoding of $\F_2$
within the group $E$, with $\enc(1)$ of order $2$ and $\add(x,y) = xy$, cannot
be realized using 4 or less group operations for $\mul$. This is left as an
exercise (one may simply explore the different possibilities).
\end{remark}


\subsection{Homomorphic encodings -- some theory}

The passage from $\SL_3(\K)$ to $\PSL_3(\K)$ motivates the search for minimal
groups with which we might work. We are going to investigate this in the case of
$\K = \F_2$.

\begin{lemma}
Assume that the group $E$ possesses a homomorphic encoding of $\F_2$, with
$\enc(0)= e$, the identity of $E$, and $\enc(1) = \sigma \in E$. Then $\sigma$
does not belong to the centre $\Zca(E)$ of $E$. It follows that the group
$E/\Zca(E)$ possesses a homomorphic encoding of $\F_2$.
\end{lemma}

\begin{proof}
  We proceed by contradiction and suppose that $\sigma \in \Zca(E)$. Since
  $\enc(\K)=\{ e, \sigma \}$ is contained in the centre of $E$, and since
  $\add$ is polynomial, there exists a function $f \colon E^2
  \longrightarrow E$ which agrees with $\add$ on $\enc(\K)$ and is of the
  form $f(x,y) = a x^n y^m$ for some integers $n, m$.
  
  We must have $f(e, e)= e$ (since $0+0=0$ in $\F_2$), hence $a = e$. Then, from
  $f(\sigma, e) = \sigma$ we get $\sigma^n = \sigma$, so we may continue the
  proof assuming $n=1$; similarly, we reduce to $m=1$, and finally $\add$
  coincides on $\enc(\K)$ with the function $f$ defined by $f(x,y) = xy$. Note
  also that $f(\sigma, \sigma) = e$ (since $1+1 = 0$ in $\F_2$), so $\sigma^2 =
  1$.
  
  Let us now turn to the multiplication. Using similar arguments, we see that
  $\mul$ coincides on $\enc(\K)$ with a map $g \colon E^2 \longrightarrow E$
  of the form $g(x,y) = b x^n y^m$ with $n, m \in \{0,1\}$ since $\sigma$ has
  order 2. From $g(e,e) = e$ we deduce $b = e$; from $g(\sigma, e) = e$ we get
  $\sigma^n = e$, so $n = 0$, and by symmetry $m = 0$. Finally, $g$ is the
  constant function equal to $e$, which is absurd, as the relation $g(\sigma,
  \sigma) = \sigma$ cannot be satisfied.
  
  The passage from $E$ to $E/\Zca(E)$ can then be carried out as described
  above.
  \end{proof}

\begin{corollary}
  If $E$ is abelian, then it does not possess a homomorphic encoding of $\F_2$.
\end{corollary}
This is clear as $\Zca(E)=E$ in this case.

\begin{corollary}
  Let $p$ be a prime number and let $E$ be a $p$-group. Then $E$ does not
  possess a homomorphic encoding of $\F_2$.
\end{corollary}

\begin{proof}
  Suppose that $E$ is a group of order $p^n$ possessing a homomorphic encoding
  of $\F_2$, and assume that it is chosen with $n$ minimal. The group $E$ is
  nontrivial as it contains $\enc(\K)$, and its centre is then also nontrivial
  (a basic fact about $p$-groups). From the lemma we see that $E/\Zca(E)$ also
  has a homomorphic encoding, even though it is a $p$-group of order less than
  that of $E$. This is a contradiction.
\end{proof}

This can be generalized immediately to nilpotent groups, and we conclude that
these do not possess homomorphic encodings of $\F_2$.

It is natural at this stage to look at simple groups. In principle, this is an
excellent idea, in view of the following theorem:

\begin{theorem}[Maurer \& Rhodes \cite{maurerrhodes}]
Suppose $E$ is a finite, simple, non-abelian group. Then any map $f \colon E^n
\longrightarrow E$, for any integer $n$, is polynomial.
\end{theorem}

We deduce, of course, that such an $E$ admits many homomorphic encodings of $\K$
for any $\K$ whose cardinality does not exceed that of~$E$.

In practice, however, even though the proof given in \cite{maurerrhodes} is
constructive, and may in principle be used to express any given function $f$ in
terms of the multiplication in $E$, the corresponding expression may involve
hundreds or even thousands of multiplications, making it hardly usable in
practice.

\subsection{Return to the general discussion}
\label{subsec-subgroup-membership}
Having established that examples of
homomorphic encodings abound, we return to the general strategy. Assume we have
chosen a group $E$ with a homomorphic encoding of the ring $\K$, and we are
looking for a group $L$ with a surjective homomorphism $\pi \colon L
\longrightarrow E$. The group $L$ is to be made public, while $\pi$ is the
secret key. If we are to achieve CPA security, then the attacker will have
access to many elements of $Z=\ker(\pi)$, in fact so many that with overwhelming
probability they will have a family of generators for $Z$; and they must not be
able to predict whether a new element $x \in L$ belongs to $Z$ or not. In
computational group theory, the {\em subgroup membership problem} is, precisely,
that of deciding whether an element of a group belongs to a given subgroup. Thus
we want to set things up so that that the subgroup membership problem is hard.

What sort of group should we employ for $L$? By this we ask the type of machine
representation to use. One naturally turns to permutation groups when working
with groups on a computer. However, the available algorithms are so powerful
that the subgroup membership problem becomes trivial.

To an extent, the same applies to matrix groups over finite fields. And methods
of linear algebra come into play, usually allowing easy solutions for the same
problem.

Groups presented by generators and relations are more promising to us, with most
tasks being very difficult to perform. However, we still need to be able to
store group elements easily, and to multiply them efficiently. For this, we
propose the use of {\em rewriting systems}, to be discussed in the next section.

Before we turn to this, we point out that we will often, though not always, have
an ``ambient'' group $G$ that properly contains $L$. It sometimes arises
naturally, for example $G$ could be a symmetric group or a general linear group.
Moreover, it can be advantageous to be in a situation where the attacker, forced
to work within $G$, is unable to identify the elements of $L$.

That said, in some cases one prefers to work with $G = L$. One reason is that it
may be computationally more expensive to derive the ``rules'' for the rewriting
system, as in the next section, for a larger group than $L$ itself.

In terms of notation, in any case, at this point we have the following objects:

\[ \begin{CD} 
   Z @>{\subset}>> C @>{\subset}>> L @>{\subset}>> G \\
       @.      @VV{\pi}V      @VV{\pi}V   @.      \\
        @.     \enc(\K) @>{\subset}>>    E      @.     
  \end{CD}
  \]
In the discussion above we have already introduced the groups $E$ and $L$ and
the map $\pi$; here we have also put $C = \pi^{-1}(\enc(\K))$ and $Z =
\pi^{-1}(\enc(0)) = \ker \pi$. The deciphering map is $\dec = \enc^{-1}\circ
\pi$.

The next section, on rewriting systems, is written for a generic group $G$, and
this is consistent with the above, that is, in practice it will the group $G$
appearing in the diagram.

\section{Rewriting systems} \label{sec-rws}

Rewriting systems are at the heart of  our protocol, and it seems
useful to briefly present their theory. Chapter~12 of \cite{holt} is an
excellent reference.

\subsection{First definitions} \label{subsec-rws-defs}

When $A$ is a (usually finite) set, we denote by $A^*$ the set of words over $A$
(that is, finite sequences of elements of $A$, including the so-called
\emph{empty word}, denoted by $\varepsilon$). We will say that $A$ is the
\emph{alphabet} we are working with. Note that $A^*$ is a monoid under the
operation of concatenation; we will write $uv$ for the concatenation of the
words $u$ and $v$.

A \emph{rewriting system} over the alphabet $A$ is a finite set of \emph{rules},
i.e., pairs of words over $A$, with the suggestive notation $\lambda
\longrightarrow \rho$ for the pair $(\lambda, \rho)$. Such a rule is said to
\emph{apply} to a word of the form $u \lambda v$, giving, by definition, the
word $u \rho v$.

We write $w_1 \longrightarrow w_2$ to indicate that there exists a rule in the
system that applies to $w_1$ to produce $w_2$. The notation $w_1
\longrightarrow^* w_2$ indicates that there is a sequence of rule applications
leading from the word $w_1$ to the word $w_2$. Finally, we write $\lrs$ for the
equivalence relation generated by $\longrightarrow$; in other words, $w_1 \lrs
w_2$ means that one can go from $w_1$ to $w_2$ by applying rules \emph{possibly
in reverse}.

More pictorially, one can construct a graph whose vertices are the elements of
$A^*$, with a directed edge from $w_1$ to $w_2$ whenever $w_1 \longrightarrow
w_2$. Then $w_1 \longrightarrow^* w_2$ means that there is a directed path from
$w_1$ to $w_2$, whereas $w_1 \lrs w_2$ means that $w_1$ and $w_2$ lie in the
same connected component (of the underlying undirected graph).

Suppose we have a \rws{} over the alphabet $A = \{ x_1, \ldots, x_n \}$, and
that we are also given a monoid, which we will denote $G$ since in the
overwhelming majority of cases it will in fact be more precisely a group. Let us
choose elements $\bar x_1, \ldots, \bar x_n \in G$. A word $w \in A^*$ gives an
element $\bar w \in G$ by replacing each occurrence of $x_i$ by $\bar x_i$ and
multiplying in $G$, so that the map $A^* \longrightarrow G$ that sends $w$ to
$\bar w$ is a monoid homomorphism. We will assume that the $\bar x_i$ generate
$G$, so that this homomorphism is surjective. Note that $\bar w$ is called the
{\em evaluation} of $w$ (with respect to the choice of generators for $G$), and
we shall also say that $w$ {\em evaluates} to $\bar w$.

We say that the \rws{} is \emph{compatible} with $G$ [with respect to the
elements $\bar x_i$] if the relation $w_1 \lrs w_2$ is equivalent to $\bar w_1 =
\bar w_2$. In particular, for every rule $\lambda \longrightarrow \rho$, we must
have $\bar \lambda = \bar \rho$, or more briefly, the “rules hold in $G$”. All
the words in the connected component of $w$, in the sense above, therefore
define the same element of $G$, and compatibility means that this induces a
bijection between the connected components and the elements of $G$.

A word $w \in A^*$ is said to be \emph{reduced} with respect to a \rws{} if no
rule applies to $w$ (one also sometimes says that $w$ {\em reduces to} $v$ to
indicate the relation $w \longrightarrow^* v$). A \rws{} is said to be
\emph{noetherian} if there is no infinite directed path in the graph defined
above, or more informally, if starting from any word and applying rules, one
always reaches a reduced word in finite time. Noetherianity is easy to guarantee
when one has chosen a total order $<$ on $A^*$, compatible with concatenation in
the obvious sense, such that for each $w$, there are only finitely many $v$ with
$v < w$: it suffices to ensure that for each rule $\lambda \longrightarrow
\rho$, we have $\rho < \lambda$.

For example, let us write $<$ for the \emph{shortlex} order on $A^*$: the
relation $v < w$ holds if $v$ is strictly shorter than $w$, or if $v$ and $w$
have the same length and $v$ comes before $w$ in the lexicographic order. In
most of our examples below, each rule $\lambda \longrightarrow \rho$ will
satisfy $\rho < \lambda$ in shortlex order. The corresponding systems will thus
be automatically noetherian, and usually we will not mention this fact.

Finally, to conclude this long list of definitions, a \rws{} is said to be
\emph{confluent} if it is noetherian and if for each word $w \in A^*$, there
exists a \emph{unique} reduced word $v$ such that $w \longrightarrow^* v$. It is
not hard to show that this additional condition is equivalent to stipulating
that each connected component of the graph contains a unique reduced word.

We will also speak, though this is not standard terminology, of a \emph{bounded}
\rws{} when there exists an integer $N$ such that every word $w$ reduces to a
word of length $\le N$. Since the alphabet is usually finite, this is equivalent
to requiring that there be only finitely many reduced words.

\begin{example} \label{ex-S3-one} On the alphabet $A = \{ a, b \}$, define the
  \rws{} with rules:
  \begin{itemize}
  \item[(R1)] $a^2 \longrightarrow \varepsilon$,
  \item[(R2)] $b^2 \longrightarrow \varepsilon$,
  \item[(R3)] $bab \longrightarrow aba$.
  \end{itemize}
  (Recall that $\varepsilon$ denotes the empty word.) One checks by inspection
  that the reduced words are $\varepsilon$, $a$, $b$, $ab$, $ba$, $aba$.
  
  Let $G = S_3$, the group of permutations of $\{1, 2, 3\}$. Set $\bar a =
  (1,2)$ and $\bar b = (2,3)$. The relations $\bar a^2 = e$, $\bar b^2 = e$, and
  $\bar b\bar a\bar b = \bar a\bar b\bar a$ hold in $G$ (where $e$ is the
  identity element of $G$); in the notation above, it follows that $w_1 \lrs
  w_2$ implies $\bar w_1 = \bar w_2$.
  
  Moreover, the 6 reduced words above, when evaluated in $G$, yield the 6
  elements of the group. This shows that these words belong to distinct
  connected components. In other words, the \rws{} is confluent. At the same
  time, we have established that the system is compatible with $G$. \finex
\end{example}
  
\begin{example} \label{ex-S3-two} To make a variant, note the relation $(\bar
  a \bar b)^3 = e$ in $G$. This suggests replacing (R3) with:
  \begin{itemize}
    \item[(R3')] $(ab)^3 \longrightarrow \varepsilon$.
  \end{itemize}
    
  The resulting \rws{} is still compatible with $G$ -- this is rarely the
  problematic point, and we leave it as an exercise. (The key point is that
  the three relations $\lambda = \rho$, where $\lambda \longrightarrow \rho$
  is one of the three rules, constitute a monoid presentation for $G$. That
  this implies compatibility is a relatively straightforward general fact.)
    
  For now, we want to emphasize the differences between the two systems, and
  in particular the fact that the second example is not confluent. Indeed, the
  reduced words are now $\varepsilon$, $a$, $b$, $ab$, $ba$, $aba$, $bab$,
  $abab$, $baba$, $ababa$, $babab$, $bababa$, so there are 12 of them, whereas
  compatibility with $G$ shows that there are only 6 equivalence classes. On
  the other hand, the system is still bounded. \finex
\end{example}

\begin{example} \label{ex-S3-three} Still with the group $G = S_3$, one
  might be tempted to introduce $\bar r = \bar a \bar b$, which satisfies
  $\bar a \, \bar r^2 = \bar r \, \bar a$, as well as the \rws{} defined
  over the alphabet $\{ a, r \}$ by the rules
  \begin{itemize}
    \item[(R1)] $a^2 \longrightarrow \varepsilon$,
    \item[(R2)] $r^3 \longrightarrow \varepsilon$,
    \item[(R3)] $ar^2 \longrightarrow ra$.
  \end{itemize}
  This system is still compatible with $G$.  A remarkable feature is that
  the words $(ar)^n$, for $n \ge 1$, are all reduced: there are infinitely
  many reduced words, and the system is not even bounded (let alone
  confluent).\finex
\end{example}



A major problem is that of finding, given a group or monoid $G$, a bounded
\rws{} which is compatible with $G$. Before reviewing the possible approaches,
let us cite a result from \cite{sims1994computation} (see Proposition 2.7 in
particular): 

\begin{theorem} \label{thm-uniqueness-rws} Let $G$ be a monoid, let $\bar x_1,
\ldots, \bar x_n \in G$ be generators, and let us use the alphabet $A= \{ x_1,
\ldots, x_n \}$. Fix an appropriate order $<$ on $A^*$. Then there exists a
unique reduced, confluent \rws{} on this alphabet which is compatible with $G$,
and such that each rule $\lambda \longrightarrow \rho$ satisfies $\rho <
\lambda$.
\end{theorem}

Here a \rws{} is called ``reduced'' when for each rule $\lambda \longrightarrow
\rho$, neither $\lambda$ nor $\rho$ contains the left hand side of {\em another}
rule. Also note that this theorem allows \rwss{} with an infinite number of
rules, but the system mentioned is finite if $G$ is finitely presented (in
particular, if $G$ is finite).

This result is not constructive and does not provide a way of writing down the
rules (for this, read the next subsections). However, the uniqueness statement
is very convenient, allowing us to speak of ``the'' \rws{} defined by the choice
of generators of $G$ (for an order which is understood, usually shortlex).

\subsection{The Knuth-Bendix algorithm}

This celebrated result must be mentioned in any survey of the theory of \rwss{},
even though in practice it cannot be applied successfully in the examples of
interest to us. Thus we remain brief. In fact, we only state a particular case.

\begin{theorem}
Resume the setting of Theorem \ref{thm-uniqueness-rws}, with $G$ finite. Suppose
we have a presentation for $G$ as a monoid, given by a finite list of relations
$\lambda_i = \rho_i$, $1 \le i \le r$, where $\lambda_i$ and $\rho_i$ are words
on the alphabet $A$. Assume $\rho_i < \lambda_i$ for each index $i$, and
consider the initial \rws{} given by the rules $\lambda_i \longrightarrow
\rho_i$.

Then there exists an algorithm which terminates in finite time, taking the rules
$\lambda_i \longrightarrow \rho_i$ as input, and returning the reduced,
confluent \rws{} which is compatible with $G$.
\end{theorem}

\begin{example}
  Let us return to Example \ref{ex-S3-two}, with its three rules (R1), (R2) and
  (R3'). Certainly the relations $a^2 =\epsilon$, $b^2=\epsilon$ and
  $(ab)^3=\epsilon$ form a presentation for $S_3$ (as a monoid or as a group).
  If we feed these three rules to the Knuth-Bendix algorithm (using the shortlex
  order), it returns the three rules of Example \ref{ex-S3-one}. To give an idea
  of what the algorithm does, it will examine the left hand sides $ababab$ and
  $bb$ and make them overlap, forming the word $abababb$; the latter can be
  reduced to $ababa$ using (R2), and to $b$ using (R3'), so the algorithm
  ``discovers'' the relation $ababa = b$ and adds the rule $ababa
  \longrightarrow b$. Reducing $ababaa$ in two different ways gives the rule
  $abab \longrightarrow ba$. Finally, reducing $aabab$ in two different ways, we
  find $bab \longrightarrow aba$ as requested.
\end{example}

\begin{example}
Consider $G= S_n$, the symmetric group of degree $n$. We choose the generators
$\sigma_i = (i, i+1)$ for $1 \le i < n$. We use the very well-known presentation
given by the relations $\sigma_i^2 = \epsilon$ for all $i$, $\sigma_i \sigma_j =
\sigma_j \sigma_i$ when $|j - i| \ge 2$, and $\sigma_{i+1} \sigma_i \sigma_{i+1}
= \sigma_i \sigma_{i+1} \sigma_i$ for $1 \le i < n-1$.

Working with the shortlex order (and of course $\sigma_i < \sigma_{i+1}$), we
can use the Knuth-Bendix algorithm on the rules $\sigma_i^2 \longrightarrow
\varepsilon$, $\sigma_j \sigma_i \longrightarrow \sigma_i \sigma_j$ when $j >
i+1$, and $\sigma_{i+1} \sigma_i \sigma_{i+1} \longrightarrow \sigma_i
\sigma_{i+1} \sigma_i$.

When $n=3$, the algorithm returns the set of rules as such, and we recover the
previous example. For $n > 3$ however, there are $n^2 - 2n +1$ rules: the
unmodified $\sigma_i^2 \longrightarrow \varepsilon$, $\sigma_j \sigma_i
\longrightarrow \sigma_i \sigma_j$ when $j > i+1$, and 
\[ \sigma_i \sigma_{i-1} \cdots \sigma_{j} \sigma_i \longrightarrow \sigma_{i-1}
\sigma_i \sigma_{i-1} \cdots \sigma_j \]
whenever $j < i$. The first mention of this in print seems to be Le Chenadec's
PhD thesis \cite{lechenadec}. 
\end{example}

This example is exceptionally simple. For $n=10$, there are less than 100 rules,
while the \rws{} corresponding to a random choice of generators for $S_{10}$ has
typically several millions of rules.

\subsection{The Froidure-Pin algorithm}

This is an alternative algorithm for
computing the reduced, confluent \rws{} corresponding to a choice of generators
in a finite monoid $G$ (for shortlex, as usual), first introduced in
\cite{froidure1997algorithms}. 

It is very easy to summarize the algorithm. Suppose we have computed the list of
all reduced words of length $\le n$, as well as the set of all rules of the form
$\lambda \longrightarrow \rho$ with the length of $\lambda$ also $\le n$ (for
$n=0$ we have the empty word and no rules). We examine each reduced word $w$ of
length $n$, and form $wa$ where $a$ is a single letter; we arrange for the words
$wa$ to come out in lexicographic order. If some rule applies to $wa$, we
discard it. Otherwise, we evaluate $\overline{wa}\in G$ and check if it is of
the form $\bar u$ for some reduced word $u$ already computed; if so, we have
discovered the rule $wa \longrightarrow u$, if not, we have a new reduced word
$wa$. 
The algorithm terminates if we have not found a single new reduced word of
length $n+1$. This will eventually happen, since we assume that $G$ is finite
and there are, at any time, fewer reduced words than elements of $G$.

\begin{example}
Let us return to Example \ref{ex-S3-three} and run the Froidure-Pin algorithm
with $\bar a=(12)$ and $\bar r=(1,3,2)$ (since we have an initial set of rules,
we could equally well have run the Knuth-Bendix algorithm on them, of course,
but note that for Froidure-Pin the input is really just the generators). The
output is:
\begin{itemize}
  \item $ a^2 \longrightarrow \varepsilon $
  \item $ r^3 \longrightarrow \varepsilon $
  \item $ ara \longrightarrow r^2 $
  \item $ ar^2 \longrightarrow ra $
  \item $ rar \longrightarrow a $
  \item $ r^2a \longrightarrow ar $.
\end{itemize}
\finex
\end{example}

For our purposes, it is of the utmost importance to be able to work with {\em
random} generators, and compute the corresponding rules. In this case, the
Froidure-Pin algorithm tends to give better results than the Knuth-Bendix
algorithm.

Moreover, we can decide to stop the Froidure-Pin algorithm at any point, in
particular when the \rws{} appears to be bounded -- perhaps in some weak,
probabilistic sense, which is checked empirically.

\begin{example}
We pick the following generators for $S_9$ randomly, using GAP: $\bar a=
(1,5)(2,4,8,7,9,3,6)$ and $\bar b= (1,7,9,3)(2,5,6)$. The Froidure-Pin algorithm
computes very rapidly the $104,110$ rules which constitute the complete \rws{}.
The longest left hand side of a rule is of length 22.

However, with only $6,325$ rules, of length no more than 17, we obtain a \rws{}
which appears bounded for all practical purposes (we comment on this phrase
below).
\end{example}

Finally, it is possible to mix the two approaches, letting Froidure-Pin run for
a while, and then applying Knuth-Bendix to get the remaining rules.

\subsection{The use of rewriting systems for homomorphic encryption; pseudo-boundedness} \label{subsec-recap}

We want to bridge explicitly the material from the previous section with the
present considerations.

Suppose we have picked a group $G$, with a subgroup $L$ as in the previous
section, itself equipped with a homomorphism $\pi \colon L \longrightarrow E$.
The idea is to pick random generators for $G$, and to compute a corresponding
\rws{}. The latter is made public. If we let $\Gc$ denote the set of reduced
words, with composition law ``concatenate-then-reduce'', then we obtain a magma
with a homomorphism $\varphi \colon \Gc \longrightarrow G$.

When the \rws{} is confluent, the magma $\Gc$ is a group and $\varphi$ is an
isomorphism. However, if we stop the Froidure-Pin algorithm before it is
completed, then $\Gc$ is certainly not a group (the composition law which we
have defined is not associative).

We will not insist on mentioning $\Gc$. It seems much simpler, in practice, to
deal with words and keep in mind that they represent elements of $G$. At any
point, we may decide to reduce a word using the rules, typically because we want
to shorten it, and this does not change the underlying element of $G$ being
represented. (Every now and then we may abuse the language and speak, say, of an
element of $L$ to mean a word which represents an element of $L$; we do this below.)

The attacker sees many words which represent elements of $Z$, or of
$C\smallsetminus Z$, in the notation of \S\ref{subsec-subgroup-membership}.
Their task is to predict, given a new word which is known to represent an
element of $C$, whether it represents an element of $Z$ (breaking the scheme is
obviously equivalent to being able to identify the ciphers of $0$).

We can now expand on what we mean by a \rws{} which ``appears bounded for all
practical purposes''. The reason we need the rules at all is that we want to
store the elements of $G$ with a finite, indeed bounded, amount of memory, so
each element of $G$ should be representable by a word whose length is less than
some absolute constant. However, we can relax this a little bit. In practice, we
are going to see elements of $L$, and only their images under $\pi$ are of
importance, and we may multiply them by elements of $Z = \ker\pi$ without
changing their decryption. Very often, if we do this repeatedly and reduce the
result with the rules every time, we can find very short words for the ciphers
{\em even if the \rws{} is not strictly speaking bounded}. In practice, we check
that the reduced words appear to be of bounded length by performing some
reductions and concatenations at random, and this works well (see Appendix
\ref{sec-appendix} for implementation details). We say that a \rws{} is {\em
pseudo-bounded} when it passes our test. The process of finding a shorter word
using multiplications by elements of $Z$ we call {\em randomized reduction}.

\section{The protocol} \label{sec-protocol}

We present our entire protocol, for the homomorphic encryption of the elements
of a commutative ring $\K$, in the standard fashion. Note that, as presented,
the scheme allows many variants; to clarify things, we present a possible choice
at every step, with some parameters kept generic. This is then illustrated in
two ways. First, we provide a toy example, with concrete values and all
computations done. The latter is of course too small to be secure, but we think
its simplicity will help the reader understand the general case. Second, we have
exposed many implementation details in Appendix \ref{sec-appendix}.

\subsection{The complete protocol (private key version)}

\subsubsection{Encoding} We pick a group $E$ and a homomorphic encoding of $\K$
within $E$.

\runningexample{We proceed as in Example \ref{ex-S6}, which we reproduce here
for convenience. We have $\K = \F_2$, $E= S_6$, we define $\enc(0)= Id$ and
$\enc(1)=(15)(34)$. We put $\add(x,y) = xy$ and $\mul(x,y) =  a_1x a_1 a_2y a_2
a_1x a_1 a_2y a_2$ where $ a_1=(12)(56)$ and $ a_2=(35)$.}

\subsubsection{Key generation} We choose a group $G$ containing a subgroup $L$
which is endowed with a group homomorphism $\pi \colon L \longrightarrow E$. We
pick an integer $d$ and a set of $d$ generators $\bar x_1, \ldots,\bar x_d$ for
$G$; these generators constitute the private key. Next, we compute the rules of
the corresponding rewriting system, for example using the Froidure-Pin
algorithm.

\runningexample{Resuming the example, we pick $G= S_n$ for some $n \ge 6$, and
$L= S_6 \times S_{n-6}$ (here $S_{n-6}$ is seen as acting on $\{ 7, 8, \ldots, n
\}$). The map $\pi$ projects onto the first factor.}

\subsubsection{Encryption} \label{subsubsec-encrypt} To encode $k \in \K$, pick
an element $x \in L$ such that $\pi(x) = \enc(k)$. Then solve the word problem
in $G$ is order to write $x$ as a product $x= \bar x_{i_1} \cdots \bar x_{i_k}$.
Construct the word $w= x_{i_1} \cdots x_{i_k}$. Optionally, reduce $w$ using the
rewriting rules. The word $w$ is the encoding of $x$. 

\runningexample{In the current example, put $x_1= (15)(34) \in G$. To
obtain $x$ as above, we pick any $z \in Z = \{ 1 \}  \times S_{n-6}$ uniformly
at random and put $x= zx_1$. Note that the word problem is easily solved in a
permutation group, using software such as GAP. Please also see Appendix
\ref{sec-appendix} for improvements.}

\smallskip We note that the functions $\add$ and $\mul$, which are part of the
encoding, involve some constants taken from $G$ -- in the discussion after
Definition \ref{def-hom-encoding}, these are called $ a_i$. It is necessary for
the homomorphic operations (see below) to compute, exactly as above, an element
$\psm a_i \in L$ such that $\pi(\psm a_i) =  a_i$, and to write it as a word
$w_i$ in the generators. This word may optionally be reduced, and is made
public.

\runningexample{In our example we have just $ a_1 = (12)(56)$ and $ a_2 = (35)$, we can take $\psm a_i = a_i$ since $E$ is identified with a subgroup of $L$ here, and two words $w_1$ and $w_2$ are computed and made public.}

\subsubsection{Decryption} To decrypt the word $x_{i_1} \cdots x_{i_k}$, compute
the product $g=\bar x_{i_1} \cdots \bar x_{i_k}$ in $G$. Things have been
arranged so that this is an element of $L$, to which $\pi$ may be applied. Then
$\pi(g) = \enc(k)$ for some $k \in \K$, and the original word decrypts to $k$.
We write $\dec(w)$ for the decryption of the word $w$.

\runningexample{Back to our example again, the element $g$ belongs to $S_6
\times S_{n-6}$, and its decryption is $0$ if it fixes the points $1, 2, 3, 4,
5,6$, while it is 1 if its restriction to this set of six points gives the
permutation $(15)(34)$. In the end we see that it is enough to compute $g(1)$
which is 1 or 5 according as the decryption of the message is 0 or 1.
}

\subsubsection{Homomorphic operations} The functions $\add$ and $\mul$ are
polynomial, so are given by formulae which can be lifted to formulae on words,
with $w_i$ replacing $ a_i$, and the basic operation being concatenation. This
defines two functions $\smb\add$ and $\smb\mul$. When $c_1$ and $c_2$ are two
ciphers, with $\Dec(c_1) = k$ and $\Dec(c_2) = \ell$, then $c_3=\smb\add(c_1,
c_2)$ is another cipher with $\Dec(c_3) = k+\ell$. Likewise $\Dec(c_4) = k \cdot
\ell$ when $c_4= \smb\mul(c_1,c_2)$. Optionally, we may reduce $c_3$ and $c_4$
using the rules.

\runningexample{In the running example the operations on words are
$\smb\add(x,y)= xy$ and $\smb\mul(x,y) =  w_1 xw_1w_2yw_2w_1xw_1w_2yw_2$. }

\subsection{A toy example} \label{subsec-toy} We describe here a toy example, which is not secure
but easily discussed. We take $n=9$ and $d=8$. We pick the following 8 elements
of $G=S_9$ at random: $\bar a= (1,7,4,2,6)(3,5,9,8)$, $\bar b= (1,2,3)(5,6,7,9)$,
$\bar c= (1,3)(2,8)(4,9,6,7,5)$, $\bar d= (1,8,3,5,7,6,2,9,4)$, $\bar e=
(1,3,6,2,8,4,5,7)$, $\bar f= (1,2,6,4,9,8,5,7)$, $\bar g=
(1,5)(2,9,4,7)(3,6,8)$, $\bar h= (1,3,7,4,8,6,9)$.

We check that they do generate $G$. Then we compute the rules. The {\em
complete} set of rules, which we can obtain on this rather small example, has
976,242 rules such as \begin{itemize}
\item $bcbdefa 	\longrightarrow	 dbgbb$, 
\item $gchfhcd 	\longrightarrow	 adbagcg$, 
\item $habhba 		\longrightarrow	 eddcdb$, 
\item $haghbfe 	\longrightarrow	 bdbghgg$, 
\item $aacecch 	\longrightarrow	 gghhhb$, 
\item $gahbggh 	\longrightarrow	 ddebbf$, 
\item $debghcf 	\longrightarrow	 fgea$,
\end{itemize}

\noindent and so on.

As will be clarified below (see \S\ref{subsec-brute}), there are just $(n-6)! =
3! = 6$ ciphers for each message. In this case they are 
\[ \{ \varepsilon,~ eeffhaf,~ ddgdfa,~ afedg,~ afcfgbf,~ bafdaf \}
\]
for the message 0 and 
\[ \{ aehbfcf,~ dhcfed,~ adhcbc,~ cachbf,~ fhabhe,~ dfbbc \}
\]
for the message 1.

We can alternatively decide to stop computing the rules when the system appears
pseudo-bounded (as in \S\ref{subsec-recap}). This happens already with 118,451
rules. In this case we can produce hundreds of ciphers of 0, for example
$baabebcfaafhc$, $ aaahhfdbhf$, $ afbfhcagbhhbebfbgaaecff$, $ bhbfbebggehe$, $
cedhadfbhcfcdbcfca$, $ bagfhabdeae$, $ afbefcahhbb$, $ ddad\-bddbcfg$, $
gefcefbcbcd$, $ afcadecbdbebd$, $ bagfhabdeae$, $ ddbaghdaddgbbag$, $
bbbchbcbe$, $ aadeggfaadceb$, $ edcafdcdhhggfhfha$, $ cbecahcfdc$, $
bcagcgah\-hefb$, $ eeefbbebghc$, $ ffffffffdcbahadeg$, $
bbacdbbdgecdebcbefdbbcfdb\-gce$, $ fabdgefagdhcd$, $ bgccaaeebcdd$, $
afdcfdahaha$, $ cgdadhbc$, $ cfabcfbf\-dbbbfbedgfb$,  $ bffbacdhgdgchec$, $
bdfbgfafbbg$, $ bddcfaheedgec$, $ bbbcbb\-cbbcbbchbcbe$, $ aebefhbdeghe$, $
bfaaegahcc$, $ bbbdhbbhhahfe$, $ bbagfbhaece\-daeg$, $ bcbabaefegchc$, $
dbcgbgecechc$,  $ cbcfbdedega$, $cadebbebbca$, $ agcgbh\-gedbdbdbd$, \dots

One way to produce such extra ciphers is to start with an initial batch,
obtained as described above, and to multiply them together, optionally reducing
them. This is similar to the ``public-key variant'' discussed below.


A word about the size of messages: a random word of length 10,000 can be reduced to one of length 12, on average.

\subsection{Public-key variant}

As with any homomorphic encryption scheme, it is
easy to turn a private-key protocol into a public-key protocol. One solution
(but the reader may envisage many variants) is as follows. Say $\K = \F_p$, the
field with $p$ elements, where $p$ is a prime. First, we construct a large
database $D$ of ciphers of $0$ and make it public. To create a new cipher of
$0$, pick a collection of elements of $D$ at random and apply to them the
operation $\smb\add$ repeatedly; then, reduce the resulting word.

We also compute a single cipher of $1$, say $w$, and make it public. To create a
cipher of $k$ mod $p$, apply $\smb\add(w, -)$ repeatedly $(k-1)$ times, and call
the result $w_k$. Then compute a cipher of $0$, says $w_0$, as above. Then the
word $w_kw_0$ is a random cipher of $k$, and again we reduce it
using the rules. We point out that the reduction process efficiently obfuscates
the way a word was constructed.

\begin{remark}
Here the reductions mentioned are mandatory, for security reasons. Indeed, a
word which is a concatenation of public ciphers can be decrypted immediately,
while the reduction process will obfuscate the origin of the ciphers thus obtained.
\end{remark}

\section{Interlude: summary of notation}

The main objects of interest in the paper appear in the next diagram:

\[ \begin{CD} 
  @.       @.       @.   \Gc \\
     @.        @.    @.   @VVV \\  
  @.       @.       @.   \tilde G \\
     @.        @.    @.   @VV{SK}V \\  
   Z @>{\subset}>> C @>{\subset}>> L @>{\subset}>> G \\
       @.      @VV{\pi}V      @VV{\pi}V   @.      \\
        @.     \enc(\K) @>{\subset}>>    E      @.     
  \end{CD}
  \]
Here is a short description:

\begin{itemize}
\item $E$ is a group with a homomorphic encoding of the commutative ring $\K$.
Typically $\K = \F_2$ and $E= S_6$.
\item $G$ is a group, and $L$ is a subgroup which maps onto $E$. So far we have
suggested $G= S_n$ and $L = S_6 \times S_{n-6}$, while later sections of the
paper suggest $G= S_n\rtimes S_n$.
\item $C = \pi^{-1}(\enc(\K))$ and $Z = \pi^{-1}(\enc(0)) = \ker \pi$. The
deciphering map is $\dec = \enc^{-1}\circ \pi$.
\item $G$ has a rewriting system, with alphabet $A = \{ x_1, \ldots, x_d \}$ and
set of rules $R$, corresponding to generators $\bar x_1, \ldots, \bar x_d$ for
$G$. We put $\tilde G = \langle A | R \rangle$, the group generated by
generators and relations by $A$ and $R$ (this is the first place where $\tilde
G$ is introduced in the paper). 
\item $\Gc$ is the set of reduced words with the concatenate-then-reduce law. As
we will see, some of the attacks will be based on viewing $\Gc$ as an object
existing independently. It is not a group in general, but merely a magma. There
is a homomorphism $\Gc \longrightarrow \tilde G$ taking $x_i$ to the element
with the same name in $\tilde G$. 
\item The map from $\tilde G$ to $G$, taking $x_i$ to $\bar x_i$, is called $SK$
where the letters are suggestive of ``Secret Key''. Formally, we have defined
the tuple $(\bar x_1, \ldots, \bar x_d)$ as the secret key; hopefully it is
clear how the secret key and the homomorphism $SK$ determine one another
uniquely, and are essentially two facets of the same thing.
\end{itemize}

\section{Security evidence: analysis} \label{sec-attacks}

In this section and the next, we provide strong evidence for the security of
GRAFHEN. We start by reviewing as many attacks as possible in
\S\ref{sec-attacks}, having consulted with experts in computational group
theory, notably Florent Hivert and James Mitchell. We can fend off many of these
by simply choosing our parameters appropriately. To establish immunity against
the other attacks, some enhancements must be brought to our basic procedure, and we turn to this in \S\ref{sec-enhancements}.

Although all the necessary notation was given in compact form in the previous
section, which the reader can use as a reference, here is a more leisurely recap
with extra comments. Of course we have our group $G$; we also have an alphabet
$A$, and a \rws{}, that is a set of rules $R$ for the words on $A$, which is
compatible with $G$. When useful, we may write $\Gc$ for the set of reduced
words, with the concatenate-then-reduce law.

The attacker sees words and rules. Thus they have access, in principle, to the
group $\tilde G$ defined by generators and relations as $\tilde G = \langle A |
R \rangle$. The secret key, consisting of the generators of $G$ which the
alphabet names, can be interpreted as a homomorphism $SK \colon \tilde G
\longrightarrow G$. The latter is very often an isomorphism (for example, when
the rules are computed with the Froidure-Pin algorithm which we keep running for
long enough), but not necessarily, as we suggest below (for clarity, we point
out that it is an easy exercise to show that $SK$ is an isomorphism just when
the \rws{} is ``compatible with $G$'' as defined in \S\ref{subsec-rws-defs},
which we often assume, but in the next section we open the door to systems with
fewer rules).

On the other hand there is a surjective map $\Gc \longrightarrow
\tilde G$; this may also be an isomorphism when the system is confluent, but
this is not typical, and in general $\Gc$ is not a group or even a monoid (its
law may not be associative).

As pointed out in \S\ref{subsec-subgroup-membership}, the security of our scheme
is based on the hardness of the subgroup membership problem in a group presented
by generators and relations, namely $\tilde G$ (the attacker must decide whether
a given word belongs to the subgroup mapping to $Z$ under $SK$, of which they
have seen many members). This is undecidable in general: if the subgroup
membership problem could be decided, the particular case of the trivial subgroup
would show that the word problem is decidable in such groups, which it is not,
see \cite{novikov}. There are even groups with decidable word problem and
undecidable subgroup membership problem, see \cite{mihailovna}. Of course, the
attacker is not required to solve the problem in general but in the particular
case of $\tilde G$.

We distinguish two types of attacks. By an attack {\em on the representation} we
mean, broadly, any attack which aims at mapping $\tilde G$ into a permutation
group $P$ (or some other type of group with solvable subgroup membership
problem) in such a way that the ciphers can be told apart in $P$. Thus this
includes any attack which leads to the discovery of the secret key itself,
namely the generators of $G$ giving access to $SK$. For this reason, we may
speak informally of an attack {\em on the key} instead of an attack ``on the
representation''.

The other category is that of attacks {\em on the ciphers}, by which we mean any
attack that will lead to deciphering some messages, but without providing a way
of trivially deciphering all of them.

We start with attacks on the representation.

\subsection{Brute-force attack}  \label{subsec-brute}

We want to estimate the effort required for a
brute-force attack on the key itself. Recall that we have an alphabet, say $A =
\{ x_1, x_2, \ldots , x_d \}$, and the key consists of corresponding elements
$\bar x_1, \bar x_2, \ldots, \bar x_d$ of a (known) group $G$, which are
generators. We wish to count the number of keys, and first we must agree on what
we consider ``truly different'' keys.

Suppose $K=(\bar x_1, \ldots \bar x_d)$ is a key, and that $R_K$ is the
corresponding full set of rules (as in Theorem \ref{thm-uniqueness-rws}, and we
reserve the letter $R$ to mean the published rules); suppose $K' = (\bar x_1',
\ldots, \bar x_d')$ is another key, with set of rules $R_{K'}$. We call the two
keys {\em equivalent} when $R_K=R_{K'}$. Note that the attacker only has access
to the set of rules (or rather, in practice, to a subset of it), and has no way
of telling two equivalent keys apart. On the other hand, observe:

\begin{lemma} \label{lem-equiv-keys}
Assume that the subgroup membership problem has an easy solution in the finite
group $G$ (which is the case when $G$ is a permutation group for example). Also
assume that the homomorphism $SK$ above realizes an isomorphism $\tilde G \cong
G$. If we use a certain key for encryption, and the attacker has access to an
equivalent key, then they can decrypt all messages.
\end{lemma}

\begin{proof}
The key obtained by the attacker defines, using evaluation of words, a
homomorphism $\varphi \colon A^* \longrightarrow G$. The latter induces an
homomorphism $\tilde G \longrightarrow G$ also written $\varphi$; indeed $\tilde
G$ is built using the public rules from the set $R$, and these are valid rules
for the attacker's key. Then $\phi$ is onto and so must be an isomorphism
because $\tilde G$ and $G$ have the same order by assumption.

A word $w$ is a cipher of $0$ just when $SK(w) \in Z$, that is $w \in
SK^{-1}(Z)$. This happens precisely when $\varphi(w) \in Z' := \varphi \circ
SK^{-1}(Z)$. 

The attacker has access to many ciphers of $0$, providing generators for $Z'$
upon applying $\varphi$. Thus the condition $\varphi(w)\in Z'$ may be checked
by solving the subgroup membership problem in $G$.
\end{proof}

Naturally, if the attacker should use a key which is not equivalent to the one
employed for encryption, then the rules of the \rws{} could not be applied, lest
they should give absurd results. All in all, it seems reasonable to consider two
keys as ``truly different'' just when they are not equivalent.

Then we note the following.

\begin{lemma} \label{lem-size-brute-force}
Assume $SK$ is an isomorphism. The number of equivalence classes of keys is
approximately
\[ \frac{|G|^d}{|Aut(G)|} \, . \]

\end{lemma}

See the proof for the sense in which this is an approximation.

\begin{proof}
The group $Aut(G)$ acts on the set of keys, that is, the set of $d$-tuples of
elements of $G$ generating $G$; the automorphism $\alpha\in Aut(G)$ takes $(\bar
x_1, \ldots, \bar x_d)$ to $(\alpha(\bar x_1), \ldots, \alpha(\bar x_d) )$. 

It is clear that two keys in the same orbit are equivalent, for each relation
(or rule) satisfied by one tuple will be satisfied by the other. Conversely, when
two keys are equivalent, we may resume the notation from the previous proof to
see that they are related by the action of $\alpha = \varphi \circ SK^{-1}$.

Thus the number of equivalence classes is precisely the number of orbits of
$Aut(G)$ on the set of keys. We note that the action is free, for the condition
$\alpha(\bar x_i) = \bar x_i$ for all indices $i$ implies that $\alpha$ is the
identity (as the $\bar x_i$'s generate $G$). Thus each orbit is of size
$|Aut(G)|$.

As a result, the number of orbits is $r / |Aut(G)$ where $r$ is the number of
keys. Now, when $d$ is large enough, a random $d$-tuple of elements of $G$ will
generate $G$ with overwhelming probability, so that the number of keys is
roughly the size of $G^d$.
\end{proof}

\begin{remark}
Here are some additional comments on the probabilistic argument at the end of
this proof. If $M$ is a maximal (proper) subgroup of $G$, then the probability
that a random $d$-tuple of elements of $G$ belongs to $M^d$ is 
\[ \frac{|M^d|}{|G^d|} = \left( \frac{|M|}{|G|} \right)^d \le \frac{1}{2^d} \, .\]
Thus if $G$ possesses $m$ distinct maximal subgroups, the probability that a
random $d$-tuple of elements of $G$ does not generate $G$ (in other words, is
contained in at least one maximal subgroup) is certainly less than $m/2^d$. 

This is a very crude estimate, and the probability is often much lower. For
example, the probability that two random elements of $S_n$ generate $S_n$ or
$A_n$ tends to $1$ rapidly as $n$ increases
(cf. \cite{dixon1969probability}, \cite{eberhard2019probability}).

\end{remark}

\begin{remark}
We have relied a couple of times on the assumption that $SK$ be an isomorphism.
This is extremely likely to be the case in practice, as the definition of
$\tilde G$ exposes millions of rules satisfied by the generators of $G$. In the
unlikely event that $SK$ should fail to be an isomorphism, however, the security
would be increased. Indeed, Lemma \ref{lem-equiv-keys} could potentially fail,
so that an attacker in the possession of a key which is equivalent to the secret
key might still be unable to decipher the messages.
\end{remark}

\begin{example}
In the example suggested in the previous section, with $G= S_n$ with $n > 6$, we
have $Aut(S_n) \cong S_n$, so that the number of truly different keys is
$n!^{d-1}$.

Let us describe a little more concretely what the brute force attack might look
like. The attacker is looking for generators of $G$, and these must satisfy the
rules; that is, for each rule $\lambda \longrightarrow \rho$, we must have $\bar
\lambda = \bar \rho$, where the evaluation of words is with respect to putative
elements $\bar x_1, \ldots, \bar x_d$. This is really the only condition, for
suppose conversely that generators of $G$ have been found, satisfying the rules.
Even if we assume that only a subset of the confluent \rws{} was made public,
the remaining rules could be in theory recovered (assuming $SK$ is an
isomorphism) using the Knuth-Bendix algorithm, and from the way this algorithm
works, we see that the attacker's generators would also satisfy the ``missing''
rules. From the uniqueness statement of Theorem \ref{thm-uniqueness-rws}, we see
that the rules defined by the proposed generators are just the rules defined by
the key, or in other words, the attacker has found an equivalent key.

Let us see how the attacker can take more concretely advantage of the fact that
the original key itself is not needed. We may imagine a situation as follows:
the group is $G= S_{13}$, and the attacker sees that the order of $\bar x_1$
must be 7 (for example they might see the rule $x_1^7 \longrightarrow
\varepsilon$). We see that $\bar x_1$ must be a conjugate of the cycle
$(1,2,3,4,5,6,7)$. Thus we may as well assume that $\bar x_1 = (1,2,3,4,5,6,7)$,
since we know we can conjugate our generators without changing the equivalence
class of the key. That is one less generator to look for, and we see how there
remains $|G|^{d-1}$ choices to explore.

We also point out that the exploration is not completely of a ``brute force''
type, as the attacker might wish (this is only an example) to consider the order
of the generators first, then they can consider the orders of the products of
two generators, and so on.
\end{example}

Here we must point out what a drastic difference it makes when one chooses a
scheme with $L=G$. In this situation, as $\pi$ is now defined on all of $G$, the
attacker will reason that only the images $\pi(\bar x_i)$ matter, and they will
look for elements in $E$ satisfying the rules, rather than in $G$. Of course $E$
is typically much smaller than $G$. Also, although the details may depend on the
chosen encoding, it is likely that two such $d$-tuples of generators of $E$
which are related by an element of $Aut(E)$ will be equally valuable to the
attacker. Thus the size of search space is closer to $|E|^d / |Aut(E)|$. In
order to secure the system, we recommend higher values of $d$.

\subsection{Todd-Coxeter} \label{subsec-todd-coxeter}

Suppose $\Gamma$ is a finitely presented group, and
$\Lambda$ is a subgroup given by a list of generators. Assume that $\Lambda$ has
finite index in $\Gamma$. Then the Todd-Coxeter algorithm computes in finite
time the action of $\Gamma$ on $\Gamma/ \Lambda$ (a permutation is given for
each generator of $\Gamma$). See Chapter 5 in \cite{holt}.

The attacker may wish to apply this with $\Gamma = \tilde G =  \langle A | R
\rangle$, the group generated by the alphabet $A$ subject to the relations given
by the set of rules $R$.  The attacker can pick a subgroup $\Lambda$ at random,
and if Todd-Coxeter completes in a reasonable time, they get a homomorphism from
$\Gamma$ into a symmetric group. With any luck, this homomorphism is injective
(this is very likely to happen with a group such as $G=S_n$ which possesses so
very few normal subgroups, given that one often has $\tilde G \cong G$). In this
case, the attacker has a representation of $\tilde G$ as a permutation group, in
which the subgroup membership problem is easy to solve, and the scheme is
broken.

It is notoriously difficult to estimate the complexity of the Todd-Coxeter
algorithm. At the very least, we can say that the bottleneck is usually the
amount of necessary memory, which can become enormous. Also, the greater the
order of $\Gamma$, and the greater the index of $\Lambda$ in $\Gamma$, the more
difficult it will be to see Todd-Coxeter run to the end (in principle it can run
with $\Gamma$ infinite, as long as $\Lambda$ does have finite index in
$\Gamma$). Thus we protect our scheme against such attacks by choosing $G$ large
enough; we may additionally arrange the choice of generators so that it is more
difficult to exhibit a subgroup of small index. We recommend using stress tests.

The Todd-Coxeter can be used in a different way, and this is our first
exploration of an attack on the ciphers (a more technical explanation will be
given in Appendix \ref{appendix-james}).  Modern implementations of the algorithm
can stop before the calculation is complete, and produce useful intermediate
data about the action. In particular, a sort of reduction process becomes
possible: given an element of $\Gamma$ written as a word $w$, a shorter word
$w'$ can be produced, such that $w$ and $w'$ act in the same way on
$\Gamma/\Lambda$, or at least such that $w \Lambda = w' \Lambda$. If $w'$ is the
empty word, we deduce that $w \in \Lambda$. Thus if the attacker tries this with
$\Lambda= SK^{-1}(Z)$, the group generated by the ciphers of $0$, then
membership in $Z$ can be proved in some cases. This attack works well with small
groups, and again, we protect the scheme against it by choosing $G$ large
enough. Again, see Appendix \ref{appendix-james} for more.


We consider a couple of extra attacks on the representation.

\subsection{Reducing the number of generators, first method}
\label{subsec-fp-inc}

We describe two attacks which rely on an application of
the Froidure-Pin algorithm to $\Gc$.

In principle, this algorithm was designed to be applied to a monoid, and the magma
$\Gc$ might not be a monoid, having a composition law which might not be
associative. However, Froidure-Pin will run anyway, and it will produce rules
which are true for {\em some} order of evaluation. This turns out to be very
useful.

For the first variant of this, pick two letters from the alphabet, say $x_1$ and
$x_2$, and view them as elements of $\Gc$. Then run the Froidure-Pin algorithm on
$x_1$ and $x_2$, and hope to find a good number of relations between them. Any rule
$\lambda \longrightarrow \rho$ thus found is certainly a true rule in $G$ when
the words (in $\Gc$) are evaluated, and we have found relations between $\bar x_1$
and $\bar x_2$.

The attacker may then use brute force to find which pairs $(\bar x_1, \bar x_2)$ of
elements of $G$ satisfy the relations. They are, in some sense, back to the
attack of \S\ref{subsec-brute}, but crucially $d$ was lowered to $2$.

Having found candidates for $\bar x_1$ and $\bar x_2$, one may return to
Froidure-Pin with $x_1$, $x_2$ and $x_3$, where $x_3$ is another letter taken from our
alphabet. One obtains relations, looks for $\bar x_3$ by brute force, then moves on
to $x_1, x_2, x_3, x_4$, and so on.

\subsection{Reducing the number of generators, second method}
\label{subsec-2gens-red}

Here is a variant of the previous attack which can work even if the Froidure-Pin algorithm finds {\em no rules at all} between $x_1$ and $x_2$. Recall that,
given a monoid $M$ and generators $x_1, x_2 \in M$, the Froidure-Pin algorithm
stores many words on the alphabet $\{ x_1, x_2 \}$ together with their
evaluations, which are elements of $M$.

With $M= \Gc$ just as above, we end up with a collection of words using just the
two letters $x_1, x_2$, and their evaluations in $\Gc$ which are words on the
whole alphabet $A= \{ x_1, \ldots, x_d \}$. With any luck, we can find a word in
$x_1, x_2$ which evaluates to a word in $x_1, x_2, x_i$ for some other index
$i$, and the letter $x_i$ is made redundant (as $\bar x_i$ can be expressed in
terms of $\bar x_1$ and $\bar x_2$). If we have enough relations, we may find
that all the generators are redundant except for the first two, and we have
again reduced to $d=2$.

Here we may point out that a more naive approach is also possible: one can
inspect the {\em given} rules (rather than trying to produce new ones), and look
for a rule of the form $\lambda \longrightarrow \rho$ where $\lambda$ and $\rho$
are not written with the same letters (perhaps with $\rho$ using one more letter
than $\lambda$ for example).

All these attacks encourage us to produce rules with some care, as does the next
attack (on the ciphers). The ``admissible rules'' that we will restrict to are
described in the next section.

\subsection{On the number of ciphers; attack by random reduction} \label{subsec-random-redux}

Keeping our notation as above, the number of different ``lifts'' for a single
message in $E$ is the order of $Z= \ker \pi$, and the number of different
ciphers for the same message is the number of reduced words $w$ whose evaluation
is in $Z$. We may safely assume that the attacker will reduce any word for us,
so we should not count the non-reduced words here (although they are technically
valid ciphers, of course). If the \rws{} is confluent, there is just one reduced
word for each element of $G$ (and $\Gc$, $\tilde G$ and $G$ are all isomorphic).
In practice however, we work with non-confluent systems, and as the reader has
seen in \S\ref{subsec-toy}, it becomes possible to create more ciphers.

In the example suggested in \S\ref{sec-protocol}, for which $L = S_6 \times
S_{n-6}$, we have $Z = \{  1 \} \times S_{n-6}$ so that there are at least
$(n-6)!$ ciphers for each message, but possibly many more.

It is generally not hard, in concrete examples, to show that the number of
ciphers of $0$ is in fact infinite: it suffices to find (by random search) a
word $w$ which is a cipher of $0$, is reduced, has length greater than the left
hand side of any rule, and is such that $w^2$ is also reduced. In this case all
the words $w^n$ are ciphers of $0$, and none can be reduced.



Does this improve security in a serious way? The answer is ``no'' without extra
effort of our part. The attacker in this situation will use {\em randomized
reduction}: as explained in \S \ref{subsec-recap}, the idea is to combine any
cipher $c$ with a random cipher of $0$, say $z$, thus forming $c'=\smb\add(c,z)$
which is another cipher with the same decryption as $c$; then $c'$ is reduced
using the rules, and kept as a replacement if it is shorter that $c$. The
process is then repeated.

Randomized reduction reduces very efficiently the length of ciphers. In the case
where we have artificially a large numbers of ciphers for each message because
the \rws{} is not confluent, the ``extra'' ciphers are long words and will be
filtered out. To give an idea, in the example of \S \ref{subsec-toy}, in which
$Z$ has order $6$ (which is ridiculously small), the randomized reduction of a
cipher of $0$ produces the empty word after just a few tries (and of course the
attacker will recognize the empty word as a cipher of $0$).

We conclude that the order of $Z$ must be large enough to guarantee the
existence of sufficiently many ciphers for each message, each of roughly the
same length.

However, this is not the end of the story. In the next section, we explain how
one can choose the rules in a certain way, thus departing from the usual
Froidure-Pin algorithm, to prevent random reduction from breaking the cipher (it
may still be used conveniently to reduce exceedingly long ciphers, as one
occasionally encounters, but will not produce very short ones).

\subsection{Using relations between ciphers} \label{subsec-fp-rel}

The next attack is a very generic one: almost any scheme based on groups à la
Nuida may be attacked in this way. Thus it must be taken very seriously. In fact
the only assumption is that the map $\dec$ should be a homomorphism, with values
in the additive abelian group $\K$ -- more concretely, this means that when the
homomorphic encoding was chosen, the homomorphic addition $\smb\add$ was taken
to be $\smb\add(x,y) = xy$, and this is extremely typical.

Pick a cipher $y$ with $\dec(y)$ known, and let $x$ be another cipher. Now run
the Froidure-Pin algorithm on $x$ and $y$ (this is in the group or monoid at
hand, so for us it is in $\Gc$). Any relation obtained between $x$ and $y$ gives
a relation in the ring $\K$, and it may allow us to express $\dec(x)$ in terms
of $\dec(y)$.

For example, suppose $\K= \F_2$ and $y$ is a cipher of $1$. Assume we have
discovered the relation $yxy^2 = x^2y$. Then $\dec(y) + \dec(x) + 2 \dec(y) =
2\dec(x) + \dec(y)$, which simplifies to $\dec(x) = 0$, revealing the deciphered
value of $x$.

One protection against this attack is to use for $L$ a group which possesses
many different homomorphisms to $\K$: if we stick to $K= \F_2$, this would mean
picking $L$ of the form $L= \F_2^m \times H$ for some group $H$, and making sure
that the first $m$ ``coordinates'' of any cipher (of either $0$ or $1$) are
random. Thus if there exists, for each pair $(x,y)$ of ciphers, a homomorphism
from $L$ to $\F_2$ mapping the pair to any element of $\F_2^2$, it will be
impossible to tell what $\dec$ does by the above method. In other words: the
attack described is based on the hope that any homomorphism $L \longrightarrow
\F_2$ at all is strongly constrained, and we work against that.

Another way to protect the scheme is to choose the rules in such a way that it
is very hard to find any relation between $x$ and $y$. We turn to this now.

\section{Security evidence: enhancements} \label{sec-enhancements}

In the previous section, we have explained how to protect our scheme against all
attacks envisaged, with the exception of those presented in
\S\ref{subsec-fp-inc} and \S\ref{subsec-2gens-red}. In
\S\ref{subsec-fp-extended} below, we explain how to select the rewriting rules
in such a way that immunity is guaranteed. Moreover, these ``new'' rules also
offer alternative (and better) protection against the attack of
\S\ref{subsec-fp-rel}, and also against the random-reduction of
\S\ref{subsec-random-redux}; this protection is so effective that the
requirement that $Z$ be large can be largely ignored.

Moreover, in \S\ref{subsec-semidirect}, we describe a simple method to create rewriting systems for larger groups. This is important for security, as we have seen that several attacks encourage us to pick $G$ of large order and $Z$ of large index. This can now be achieved easily.

\subsection{Admissible rules} \label{subsec-fp-extended}

The long list of attacks just given provides motivation for the following
definition. A rule $\lambda \longrightarrow \rho$ will be called {\em
admissible}, with respect to the choice of an integer $k$, when:

\begin{enumerate}
\item both $\lambda$ and $\rho$ use the entire alphabet,
\item both $\lambda$ and $\rho$ are of length no less than $k$,
\item $\lambda$ and $\rho$ have no prefix and no suffix in common.
\end{enumerate}

\begin{remark}
Of course the third condition is merely here because, as the rules correspond to
identities in a group (for us), one could simplify any common prefix or suffix,
and produce a shorter, valid rule, which in turn may not satisfy the first two
conditions. 
\end{remark}

Having chosen generators $\bar x_1, \ldots, \bar x_d$ for a group $G$, and an
integer $k$, we can produce a set of corresponding admissible rules (over the
alphabet $\{ x_1, \ldots, x_d \}$ as usual), so that the resulting \rws{} is
pseudo-bounded. For this, we merely modify the Froidure-Pin algorithm so that it
rejects any rule $\lambda \longrightarrow \rho$ which is not admissible (so
$\lambda$ is considered as a reduced word in the rest of the procedure). We need
to produce more rules to reach pseudo-boundedness, but otherwise this works well
in practice.

Such a \rws{} exhibits much less of the group structure of $G$. We can
illustrate this with $\Gc$ -- recall that this is our notation for the set of
reduced words with the concatenate-then-reduce law. With admissible rules, there
are no relations at all between the elements $x_1, \ldots, x_{d-1}$ of $\Gc$
(with the last generator $x_d$ left out), and these generate a monoid isomorphic
to that of all words on the alphabet $\{ x_1, \ldots, x_{d-1} \}$. This is in
sharp contrast, of course, with the structure of the finite group $G$.

In terms of security, restricting to admissible rules solves many problems at
once. It is obvious that the attacks of \S\ref{subsec-fp-inc} and
\S\ref{subsec-2gens-red} will not work anymore. Relations between ciphers become
extremely hard to discover, and the attack of \S\ref{subsec-fp-rel} cannot be
carried out in practice. Last but not least, it turns out that the random
reduction attack of \S\ref{subsec-random-redux} is also rendered inefficient.
Perhaps surprisingly, we end up with a very large number of ciphers of $0$, even
with the order of $Z$ fairly small.

\subsection{Semidirect products} \label{subsec-semidirect}
Next we present a simple trick allowing us to reach larger groups. Obviously we
need our group $G$ to be as large as possible, to fend off the brute-force
attack and the Todd-Coxeter attack, but large groups require large numbers of
rules. 

First we present the construction in a naive way. Given a \rws{} for $G$ on the
alphabet $A$, with rules $R_1$, and another one also for $G$ on the alphabet $B$
with rules $R_2$, we try to combine them and get a larger \rws{}. We note that
the rules in $R_i$ can be interpreted as rules for words on the alphabet $A\cup
B$, for $i=1, 2$. If we want to get a bounded \rws{} with minimal effort from
this, one possibility is as follows: for $a \in A$ and $b\in B$, find a word $w$
over the alphabet $A$ such that $\bar w = \bar b \bar a \bar b^{-1} \in G$; then
consider the rule $ba \longrightarrow wb$, which involves both alphabets (and
which is motivated by the fact that $\bar b \bar a = \bar w \bar b$). Write
$R_1\rtimes R_2$ for the \rws{} on the alphabet $A\cup B$ consisting of $R_1\cup
R_2$ together with all such ``commutation rules''. It is clear that $R_1\rtimes
R_2$ is bounded if $R_1$ and $R_2$ both are; in fact any word $w$ on $A\cup B$
can be reduced to a concatenation $u v$ with $u$ resp.\ $v$ a reduced word on
$A$ resp.\ $B$.

The form of the reduced words just given makes it plausible that $R_1\rtimes
R_2$ should not be ``compatible'' with $G$ but with a much larger group;
although we see from the outset that the larger group should have a homomorphism
onto $G$ since words on $A\cup B$ can be evaluated to elements of $G$, and all
the rules in $R_1\rtimes R_2$ hold true in $G$.

The reader who is familiar with semidirect products will guess what the ``larger
group'' is immediately. Recall that, whenever the group $H$ acts (on the left)
on the group $N$ by automorphisms, we can form the semidirect product $N\rtimes
H$, which consists of all pairs $(n,h) \in N \times H$, with the following group
law: 
\[
  (n_1, h_1) (n_2, h_2) = (n_1 (h_1\cdot n_2), h_1 h_2) \, .
\]
The following is then a simple exercise in decoding the definitions:

\begin{lemma} \label{lem-R1R2isGG}
The \rws{} $R_1\rtimes R_2$ is compatible with $G\rtimes G$, where $G$ acts on
itself by conjugation.
\end{lemma}

\begin{proof}
In general there is an embedding of $N$ into $N\rtimes H$ given by $n \mapsto
(n,1)$, whose image is a normal subgroup; there is also an embedding of $H$ into
$N\rtimes H$ given by $h\mapsto (1,h)$, whose image is not normal in general.

With this in mind, for $a\in A$ consider $\tilde a = (a,1) \in G\rtimes G$, and
for $b\in B$ consider $\tilde b = (1,b)$. This allows us to evaluate any word $w$
over $A\cup B$ to give an element $\tilde w\in G\rtimes G$.

We check readily that for each rule $\lambda \longrightarrow \rho$ of
$R_1\rtimes R_2$, we have $\tilde \lambda = \tilde \rho$ in $G\rtimes G$. As a
result, when two words $w_1$ and $w_2$ satisfy $w_1 \lrs w_2$ for the \rws{}
$R_1\rtimes R_2$, we have $\tilde w_1 = \tilde w_2$.

Now suppose conversely that $\tilde w_1 = \tilde w_2$. We may as well assume
that $w_1 = u_1 v_1$ with $u_1$ a word over $A$ and $v_1$ a word over $B$;
likewise assume $w_2 = u_2 v_2$. Thus $\tilde w_1 = (\bar u_1, \bar v_1) = \tilde
w_2 = (\bar u_2, \bar v_2)$. This identity in $G\rtimes G$ implies $\bar u_1 = \bar
u_2$ and $\bar v_1 = \bar v_2$. As we assume that $R_1$ is a \rws{} which is
compatible with $G$, we have $u_1 \lrs u_2$, and similarly $v_1 \lrs v_2$. It
follows that $w_1 \lrs w_2$.
\end{proof}

So we obtain a \rws{} for the group $G\rtimes G$, which is much larger than $G$,
essentially for free. To see what we can do with it, we must notice the
following at once.

\begin{lemma}
Let $G$ act on itself by conjugation (on the left), and form the semidirect
product $G \rtimes G$. Then $G\rtimes G \cong G\times G$. In particular, there
are two distinguished homomorphisms $p$ and $f$ from $G\rtimes G$ onto $G$,
defined by $p(x,y)=y$ and $f(x,y)=xy$.
\end{lemma}

\begin{proof}
An easy proof of the first statement is obtained by considering the normal
subgroup $N = G \times \{ 1 \}$ in $G\times G$, and the ``diagonal'' group
$\Delta = \{ (g,g) ~| ~ g \in G \}$. It is clear that $N \cap \Delta = \{ 1 \}$
and that $N \Delta = G \times G$, so by a basic criterion for semidirect
products, we have $G\times G \cong N \rtimes \Delta$. Moreover $N$ and $\Delta$
are both isomorphic to $G$ and the result follows. 

For a more explicit proof, consider $G\rtimes G$ and its projection $p \colon
G\rtimes G \longrightarrow G$ defined by $p(x,y) = y$. Crucially, the map $ f
\colon G\rtimes G \longrightarrow G$ defined by $f(x,y) = xy$ is also a
homomorphism. Combining $p$ and $f$, we obtain a map $\phi \colon G\rtimes G
\longrightarrow G\times G$ given by %
\[
  (x,y) \mapsto (f(x,y), p(x,y)) = (xy, y) \, .
\]
It is clear that $\phi$ is an isomorphism, with inverse $(X,Y) \mapsto (XY^{-1},
Y)$. 
\end{proof}

The map $f$ is crucial to us. Let us describe how it works, in terms of words.
In the proof of Lemma \ref{lem-R1R2isGG}, we have written $\tilde w$ for the
evaluation in $G\rtimes G$ of a word $w$ on the alphabet $A\cup B$. Now for $a
\in A$, the map $f$ takes $\tilde a= (\bar a,1)$ to $\bar a$, and likewise for
$b\in B$ we see that $f$ takes $\tilde b=(1,\bar b)$ to $\bar b$. Let us add,
much more informally, that when one works with \rwss{} one is rapidly tempted to
forget the ``bars'' and to rely on context to distinguish $a$ from $\bar a$,
resulting in the observation that $f$ maps $a$ to $a$ and $b$ to $b$ which,
as it turns out, does not create confusion. When seeing a word on the
alphabet $A$, we may think of the corresponding element of $G$, which can be
identified with a subgroup of $G\rtimes G$ if we prefer, and likewise for words
on $B$; when a word mixes both alphabets, the element can only be interpreted in
$G\rtimes G$.

Here is how we use all this in practice. We usually start with a choice of group
$G$ for our protocol, with initial choices of $L_0$ and $\pi_0 \colon L_0
\longrightarrow E$, in obvious notation. Then we consider $G\rtimes G$ and, for
example, put $L = f^{-1} (L_0)$ and $\pi = \pi_0 \circ f \colon L
\longrightarrow E$. Thus the group $Z= \ker(\pi)$ is much larger than $Z_0 =
\ker(\pi_0)$ (in fact its size has been multiplied by $|G|$).

Alternatively, as we may wish to increase the {\em index} of $Z_0$ as well as
its size, we may pick a smaller group for $L$ (see the example that follows). 

Then we compute two sets of rules $R_1$ and $R_2$ for $G$ and combine them, as
above, into a \rws{} $R_1 \rtimes R_2$ for $G\rtimes G$.

\begin{example}
Say we start with $G= S_{11}$ and $L_0 = S_6 \times S_5$, with projection map
$\pi_0 \colon L_0 \longrightarrow S_6$. Thus $Z_0 = \ker(\pi_0)$ has order $5!=
120$ and index $11! / 5! = 332,640$. We turn to $G\rtimes G$ and consider 
\[
  L = \{ ( (z,x), (1,x^{-1})) ~|~ (z,x) \in S_6 \times S_5  \} \, .
\]
This is a subgroup of $G\rtimes G$ and it is endowed with the homomorphism $\pi
\colon L \longrightarrow E=S_6$ given by $\pi((z,x), (1,x^{-1})) = z$. The
kernel $Z$ of $\pi$ is again isomorphic to $S_5$ so its order is again $5! =
120$, but its index in $G\rtimes G$ is now $(11!)^2 / 5! = 13,277,924,352,000$
which is more than enough to protect us against an attack using Todd-Coxeter as
in \S\ref{subsec-todd-coxeter}.
\end{example}

For yet another way of choosing $L$, see \S\ref{sec-parameters} below. Many
variants can be envisaged.

\begin{remark}
Whatever the variant, the decryption process relies on the map $f$, and to
compute its values one needs the full key, consisting of all the elements $\bar
a \in G$ for $a\in A$ and also all $\bar b\in G$ for $b\in B$. The reader is
invited to contemplate how the ``obvious'' way of creating a \rws{} for $G\times
G$ from two such systems for $G$ (by requiring letters from different alphabets
to simply commute) would give rise to a protocol for which half the alphabet can
in fact be ignored. The construction above truly mixes the two keys.
\end{remark}

\begin{remark} \label{rmk-semidirect-not-better}
While the semidirect product construction is a very efficient protection against
attacks on the ciphers, it does not increase the resistance of the scheme
against attacks on the key. Indeed, if the attacker succeeds in finding the
secret key corresponding to the ``left'' copy of $G$ (on the alphabet $A$), or
if they merely can represent efficiency this copy of $G$, then one can compute
the action of the second copy of $G$ on the first by conjugation. This is
usually (depending on $G$) enough to break the second key.
\end{remark}

\subsection{Automorphisms}

The last ``enhancement'' which we present is rather a
drastic variant on our protocol. It remains a engineering challenge to work
efficiently with the automorphisms to be introduced next, but on the other hand
the level of security provided is very appealing.

The idea is to start with a group, which we are going to call $G_0$ this time,
which is equipped with a \rws{} as above, and to work with $G= \aut(G_0)$, the
group of automorphisms of $G_0$. Concretely, if we have chosen the generators
$\bar x_1, \ldots, \bar x_d$ for $G_0$ (so that the corresponding alphabet is
$\{ x_1, \ldots, x_d \}$) then an element $\alpha\in G$ can be reconstructed
from the list $\alpha(\bar x_1), \ldots, \alpha(\bar x_d)$. In turn, if we
choose a word $w_i$ with $\bar w_i = \alpha(\bar x_i)$ for each index $i$, then
we see that $\alpha$ can be stored on a computer as the finite list of words
$w_1, \ldots, w_d$.

In this representation, if we want to compute $\alpha \circ \beta$ where
$\alpha, \beta\in G$, we start with the words $w_1, \ldots, w_d$ resp.\ $w_1',
\ldots, w_d'$ representing $\alpha$ resp.\ $\beta$; we consider each $w_j'$ and
replace in it each occurrence of $x_i$  by $w_i$, that is we form $w_j'' := w_j'(w_1,
\ldots, w_d) $, in standard notation. Then the words $w_1'', \ldots,
w_d''$ represent $\alpha \circ \beta$, since 
\[
  \bar w_j'' = w_j'(\bar w_1, \ldots, \bar w_d) = w_j'(\alpha(\bar x_1), \ldots, \alpha(\bar x_d)) = \alpha (w_j') = \alpha (\beta(\bar x_j)) \, .
\]
Briefly, composition is substitution.

For example, if we start with $G_0 = S_n$ with $n>6$, then as is classical one
has $\aut(S_n) \cong S_n$, so what we have is another representation of the
symmetric group on a computer. The rest of the protocol (with the choice of $E$
and $L$ and so on) can be carried out without changes.

The great advantage of this variant is that the subgroup membership problem
seems exceedingly difficult to deal with, in a group presented as $\aut(G_0)$,
with $G_0$ as above -- in the sense that there is essentially nothing in the
literature to even get us started. We seem to benefit from an excellent
protection against all ciphers attacks. (Attacks on the representation of $G_0$
could still work, of course, and the same precautions must be taken.)

We shall not try to substantiate any security claim. For now, as this protocol
uses rather large ciphers and leads to potentially heavy computations for the
homomorphic operations, we are bound by practical considerations to postpone a
fuller investigation.

\section{Recommended parameters \& a benchmark} \label{sec-parameters}

We end the paper with a realistic example. We shall work with $E= S_{11}$, and
leave the choice of encoding open. As $S_6 \subset S_{11}$, one may choose the
encoding proposed in Example \ref{ex-S6}. It is possible to do much better with
$S_{11}$, but the exploration of efficient encodings will be the subject of a
subsequent publication.

\begin{itemize}
\item Let $G= S_{11}$. We pick 5 generators $\bar x_1, \ldots, \bar x_5$ at
random. Usually, we ensure that the subgroup generated by $\bar x_i$ and $\bar
x_j$, for any pair $(i,j)$, is all of $G$ (otherwise the attacker has access to
``easy'' subgroups of $G$ which might be fed to the Todd-Coxeter algorithm as in
\S\ref{subsec-todd-coxeter}).

\item Use our modified Froidure-Pin algorithm as in \S\ref{subsec-fp-extended}
to obtain a pseudo-bounded \rws{} on the alphabet $\{ x_1, \ldots, x_5 \}$ whose
rules are admissible. \pierre{on prend combien pour $k$ ?}\florian{C'est quoi
$k$ déjà ? Tu parles du nombre de générateurs ? Si oui, je suggère pour
l'instant de laisser $k=5$ comme dans la dernière clé générée} \pierre{non $k$
c'est la longueur minimale des règles, comme dans la définition de
"admissible"}We obtain about 20 million rules.

\item Repeat the first two steps, and obtain a second \rws{} for $G$ on the
alphabet $\{ y_1, \ldots, y_5 \}$.

\item Combine the two \rwss{} into one for $G\rtimes G \cong G\times G$, as
described in \S\ref{subsec-semidirect}.

\item Put $E= S_{11}$ and $L = E \rtimes S_8$, a subgroup of $G\rtimes G$. The
homomorphism $f \colon G\rtimes G \longrightarrow G$ described in
\S\ref{subsec-semidirect} (which maps $x_i$ and $y_i$ to ``the element with the
same name in $G$'') takes $L$ into $E$, and this defines our map $\pi$. 

\item In practice, to encrypt $e\in E$, we pick $x\in S_8$ at random and form
$(ex^{-1},x)\in L$. 
\end{itemize}


We can now describe a benchmark we have run, in order to get a first idea of the
performance we can expect. It is notoriously difficult to design a benchmark
which is truly informative, so we will do our best to explain what is being
measured. 

On the one hand, we have considered OpenFHE, a popular, high-quality library
which is a current standard. 
 From the webside at
\url{https://openfhe.org/} we quote : ``OpenFHE is an open-source project that
provides efficient extensible implementations of the leading post-quantum Fully
Homomorphic Encryption (FHE) schemes''. It is a library written in C++,
supporting all major FHE schemes, including the BGV, BFV, CKKS, DM (FHEW), and
CGGI (TFHE) schemes.

We have used the default bootstrapping method, which is GINX (aka TFHE), with a
security of 100 bits. We have homomorphically computed, for various encrypted
bits (elements of $\F_2$) the AND operation (that is, the product). 

On the other hand, we have measured the performance of our implementation of
GRAFHEN, written in Rust. The parameters are chosen exactly as suggested above.
We add that the messages are elements of $S_6 \subset E = S_{11}$, and in the
expression $(ex^{-1},x)$ just given for the ciphers, we have in fact $e = my$
with $m\in S_6$ (the message) and $y\in S_5$ taken randomly (viewing $S_6\times
S_5$ as a subgroup of $S_{11}$ as usual). The encoding is that of
Example~\ref{ex-S6}, so we are encrypting elements of $\F_2$.

We also mention that we have decided to rely only on rules $\lambda
\longrightarrow \rho$ with the length of $\rho$ strictly less than that of
$\lambda$. We need to compute more rules in order to achieve pseudo-boundedness
with this constraint, but this speeds up the reduction process. All in all, the
particular key we have constructed involves just under 40 million rules.

To give an idea of the security afforded, we rely on Lemma
\ref{lem-size-brute-force} (and Remark \ref{rmk-semidirect-not-better}): the
number of (``truly different'') keys is $11!^4$, which is about $2^{101}$.  

We have measured the (average) time needed to concatenate and reduce two words,
which were themselves reduced. This is how one performs homomorphically the
group operation of $E$ on encrypted data. As the encoding employed uses 5 such
operations for a homomorphic multiplication (see Remark \ref{rmk-five-mul}), we
have multiplied the measured time by 5 to obtain the time needed to perform the
AND operation on two bits with our implementation.

The computations were performed on a MacBook with an Apple M4 Pro CPU, with 14
cores and 48 GB of RAM. Here are the results:

\begin{itemize}
\item OpenFHE : $26.4 ms$
\item GRAFHEN :  $7.56 \mu s$
\end{itemize}

In this proof-of-concept implementation, GRAFHEN is thus already 3500 times
faster than OpenFHE. Moreover, much more is being done to optimize the core
implementation, which is built on top of an entirely novel scheme, and will benefit greatly from dedicated engineering.

\appendix

\section{A complete example} \label{sec-appendix}

We will now provide a complete description of the practical steps necessary to
produce a key and work with our protocol. We will include some code, and this
will use Sagemath: recall that Sagemath is based on the familiar Python
language, and includes many specialized packages, most notably for us the GAP
system. (For readers familiar with GAP but not Sagemath, we point out that the
{\tt libgap} object in Sagemath allows access to GAP functions.)

Please keep in mind that our goal here is pedagogy, and all code is written so
as to be easy to understand, and not, as will be amply evident, to be efficient.

\subsection{Initial choices}

One must start by picking a group $E$ with an encoding. In this appendix we
shall stick to Example \ref{ex-S6}, with $E=S_6$, and we are thus encoding a
single bit.

Next we must choose $G$. Among other choices, we recommend using the symmetric
group $G= S_n$, and this is what we will focus on in this appendix, so one must
choose an integer $n$. We recommend $n \ge 11$ for security (large values of $n$
will make it difficult to compute the rules, see below). One also needs to pick
$d$ generators for $G$, where the integer $d$ has been chosen, and we recommend
$d \ge 4$, again for security (large values of $d$ will produce too many rules).

In order to pick random generators in practice, one can proceed as follows.
Suppose we have:

\begin{code}
n= 11
d= 5
G= libgap.SymmetricGroup(n)
\end{code}

Then the following picks $d$ random generators, which together comprise our
secret key:

\begin{code}
while True:
  gens= [ G.Random() for _ in range(d) ]
  if G.Subgroup(gens) == G:
    break
\end{code}
(Keep in mind that GAP's algorithm for producing random elements of a group will
always produce the same sequence of elements, so you must take care to properly
initialize GAP's pseudorandom number generator with a crypto-secure seed.)

We will need names for the chosen elements. Assuming $d \le 26$ we may as well
use:

\begin{code}
alphabet= "abcdefghijklmnopqrstuvwxyz"[:d]
\end{code}

We will frequently need to evaluate words in this alphabet, to recover the
corresponding permutations. This can be achieved thus:

\begin{code}
evaldict= dict( (letter,g) for letter,g in 
      zip(alphabet,gens))
def wordev(w):
  return prod(evaldict[letter] for letter in w)
\end{code}
Thus {\tt wordev("ab")} will return the product of the first two elements in the list
{\tt gens}.

We must also choose $L$ and a homomorphism $\pi \colon L \longrightarrow E$.
Here we shall use $L = S_6 \times S_{n-6}$, where the homomorphism $\pi$ is the
projection onto the first factor, while its kernel $Z$ is $\{ 1 \} \times
S_{n-6}$.  

This translates to:

\begin{code}
E= SymmetricGroup(6)
L= E.DirectProduct(libgap.SymmetricGroup(n-6))
Z= L.Embedding(2).Image()
pi= L.Projection(1)
\end{code}

\subsection{Producing elements}

We need to be able to perform the following task: given $x\in E$, find $y\in L$
such that $\pi(y) = x$. Moreover, we should find a word in our alphabet which
evaluates to $y$.

There are two ways to go about this, and it is best to use both. The first is
GAP's ability to solve the word problem in a permutation group (this boils down
to the Schreier-Sims algorithm). For this, we start with:

\begin{code}
hom= G.EpimorphismFromFreeGroup()
\end{code}
Then {\tt hom.PreImagesRepresentative(g)} returns, for any $g\in G$, a word
which evaluates to the permutation $g$. Unfortunately, this word is in another
alphabet, and more importantly, it involves the inverses of the generators,
while we generally prefer to stick to powers of the original generators. Thus we
need the somewhat awkward:

\begin{code}
orders= [ int(g.Order()) for g in gens ]
def gap_word_to_string(word):
  L = [word.Subword(i,i) for i in 
      range(1, word.Length()+1)]
  s = "".join(str(letter) for letter in L)
  for i in range(len(gens)-1, -1, -1):
    name = "x"+str(i+1)
    newname = alphabet[i]
    order = orders[i]
    s = s.replace(name+"^-1", newname*(order -1))
    s = s.replace(name, newname)
  return s
\end{code}
Armed with this we can define:
\begin{code}
def solve_word_problem(g):
  gapword= hom.PreImagesRepresentative(g)
  return gap_word_to_string(gapword)
\end{code}
For example \begin{code}
solve_word_problem(wordev("a"))
\end{code} 
should return {\tt "a"}. 

Perhaps a better example is the construction of the words needed, as described
in \S\ref{subsubsec-encrypt}, in order to perform the homomorphic operations. As
already noted in that paragraph, the elements $a_1 = (12)(56)$ and $a_2=(35)$
are trivially lifted to $L$ since we may take $\psm a_i = a_i$ for $i=1,2$ (with
the notation above, $a_i$ plays the role of $x$ and $\psm a_i$ plays the role of
$y$). What requires work is to write these as words in the alphabet, and we now
have the code for this:

\begin{code}
a1= solve_word_problem(libgap.eval("(1,2)(5,6)"))
a2= solve_word_problem(libgap.eval("(3,5)"))
\end{code}

We can also use the above to produce ciphers. A cipher of 0 can be created with:

\begin{code}
c0= solve_word_problem(Z.Random())
\end{code}

A cipher of 1 can be created with:

\begin{code}
c1= solve_word_problem(
    libgap.eval("(1,5)(3,4)")*Z.Random() ) 
\end{code}
To see why the permutation $(1,5)(3,4)$ comes up, go back to
Example~\ref{ex-S6}.  

We can improve on this, however. The above method for producing ciphers of 0
will always produce the same word for a given element of $Z$, even though we
have argued in the paper that many more ciphers are available. Moreover, when
translating the GAP words into usual strings, the inverses are replaced by
powers of the generators, making it easy for an attacker to guess the orders of
the generators, and it is best to avoid this.

The idea is simple: as $E$ is small enough, if we pick elements of $L$ at random
sufficiently many times, we will end up with an element mapping to any $x\in E$
given in advance; moreover, we can draw the {\em words} at random rather than
the actual elements.

Let us start with generators for $L$. We have plenty of choices; for example we
may rely on the fact that $S_m$ is always generated by the ``Coxeter''
generators $(i,i+1)$ with $1 \le i < m$.

\begin{code}
gensL= [ libgap.eval("("+str(i)+","+str(i+1)+")")
    for i in [1..5] + [6..(n-1)] ]
\end{code}
Next we write these as words in our alphabet:

\begin{code}
wordsL= [ solve_word_problem(x) for x in gensL ]
\end{code}

Here is how we can find a random word evaluating to an element $y\in E$, which
in turn maps to the given $x\in E$:

\begin{code}
def find_word_for(x):
  while True:
    word= ""
    for _ in range(100):
      word= word+random.choice(wordsL)
    y= wordev(word)
    if x == pi.Image(y):
      return word
\end{code}

With this, one can create a cipher of 0 with:

\begin{code}
c0= find_word_for(E.One())
\end{code}
and a cipher of 1 with:
\begin{code}
c1= find_word_for(libgap.eval("(1,5)(3,4)"))
\end{code}
This will produce a plethora of different ciphers.

\subsection{Performing the homomorphic operations}

Again we follow the recipe given in Example \ref{ex-S6}. We see that homomorphic
addition is just concatenation:

\begin{code}
def hom_add(x, y):
  return x+y
\end{code}

Homomorphic multiplication is given by:
\begin{code}
a1a2= a1 + a2
def hom_mult(x, y):
  z= a1 + x + a1a2 + y + a2
  return z + z
\end{code}

\subsection{Decryption}

This is trivially achieved as follows:

\begin{code}
def decrypt(c):
  g= wordev(c)
  if g == E.One():
    return 0
  else:
    return 1
\end{code}
Please note that the function {\tt wordev} used here relies on the list {\tt
gens}, which is the secret key.

\subsection{Computing the rules}

We have indicated all the steps to produce
ciphers, operate on them, and decipher them. However, it is of the utmost
importance in practice to be able to reduce the size of words. So we need to
compute the rules of the \rws{}.

This is achieved by the Froidure-Pin algorithm, see
\cite{froidure1997algorithms}. There is an implementation available as part of
the {\tt libsemigroups} library, written in C++, which can be interfaced with
Python.

The algorithm takes as input a family of generators for a finite monoid, and
also an alphabet for naming these. Its output is a set of rules, in fact the
rules described by Theorem \ref{thm-uniqueness-rws} for the shortlex ordering,
which can be converted to a Python dictionary. For example, if called with the
generators $(1,2)$ and $(2,3)$ for the group $S_3$ (seen as a monoid), with the
alphabet {\tt "ab"}, the output would be

\begin{code}
{ "aa" : "", "bb" : "", "bab" : "aba"}
\end{code}
exactly as in Exemple \ref{ex-S3-one}.

Let us now describe how to reduce a word with respect to a set of rules. To do
so effectively, one should use automata, as described in \cite{holt} (see
\S13.1.3 there). However, here is a toy implementation that one can use:

\begin{code}
def naive_reduce(word, rules) :
  w= word
  while True:
    lhs= possible_reduction(w, rules)
    if lhs is None:
      return w
    else:
      w= w.replace(lhs, rules[lhs])
\end{code}
In turn this uses:

\begin{code}
def possible_reduction(word, rules):
  for lhs in rules:
    if word.find(lhs) > -1:
      return lhs
  return None
\end{code}

We have preferred to use a custom implementation of the Froidure-Pin algorithm.
One reason is the ability to filter the rules, for example to exclude the rules
which are too short (see \S\ref{subsec-fp-extended} for more about this).
Another reason is that we want to be able to stop the algorithm when just
``enough'' rules have been computed, and not wait for the complete set of rules
(which might not fit into the computer's memory anyway).

By ``enough'' rules, we mean for the \rws{} to be pseudo-bounded, as defined (in
loose terms) in §\ref{subsec-recap}. Very concretely, given a collection of
rules, we perform the following test:

\begin{itemize}
  \item pick 10 random words in the given alphabet, each of length 10,000 ;
  \item reduce these words using the rules ;
  \item compute the average length $\ell$ after this reduction has been
  performed ;
  \item concatenate the 10 reduced words to form $w$, and reduce again $w$ using
  the rules, calling the result $w'$.
\end{itemize}

When this is done, if the length of $w'$ is less than $3 \ell$, we consider the
\rws{} to be pseudo-bounded. Obviously one can try different constants here (the
number 10, the length 10,000, the factor 3).

\section{Computational attacks, by James Mitchell (University of St-Andrews)} \label{appendix-james}






\newcommand{\GAP}{\textsc{GAP}~\cite{GAP4}\xspace}
\newcommand{\libsemigroups}{\textsc{libsemigroups}~\cite{libsemigroups}\xspace}
\newcommand{\libsemigroupspybind}{\textsc{libsemigroupspybind11}~\cite{pybind}\xspace}
\newcommand{\ACE}{\textsc{ACE}~\cite{HavasRamsay1995}\xspace}

\newcommand{\genset}[1]{\langle#1\rangle}


In this appendix we discuss possible attacks on the ciphers related to the
finitely presented groups and monoids generated according to the scheme in the
paper. We only consider those attacks that permit the decryption of messages
encoded using the scheme, not attacks on the key itself, which appears to be a
significantly harder problem. 


The input used in the attacks described in this section are:
\begin{enumerate}
  \item a finite monoid presentation $\mathcal{P}=\langle A|R\rangle$ (possibly
    defining a group) defining a monoid $M$;
  \item a set of words $Z$ that encode $0$;
  \item a tuple of challenge words $S=(S(0), S(1), \ldots, S(n))$ encoding $0$ and $1$;
  \item a tuple of solutions $T\in \{0, 1\}^ {|S|}$ such that the $i$th term $T(i) = 0$
    if and only if $S(i)$ belongs to $Z$.
\end{enumerate}
The aim of the attacks is to determine which of the words in $S$ correspond to
$0$ (those belonging to the subgroup $\genset{Z}$ generated by $Z$) and which
correspond to $1$ (those not belonging to $\genset{Z}$).

The main tool used for the attacks is the Todd-Coxeter algorithm for semigroups
and monoids~\cite{Coleman2024} (numbers in brackets now refer to the additional
bibliography which follows). This algorithm is implemented in the C++ library
\libsemigroups and was used through the Python bindings~\libsemigroupspybind. 

In what follows we will use the language of monoids, it might be worth noting
that the notion of 1-sided congruences for monoids generalises the notation of
subgroups for groups. The Todd-Coxeter algorithm attempts to determine the
action of the monoid $M$ on the least 1-sided congruence $\rho$ on $M$
containing $Z\times 1$. There are other algorithms for computing finitely
presented monoids (namely the Knuth-Bendix algorithm~\cite{KnuthBendix1970},
the low index congruence algorithm~\cite{Merkouri2023}, and several specialised
algorithms such as those for the so-called \textit{small overlap monoids}). The
Knuth-Bendix and the low index congruence algorithms can, in theory at least,
be applied to the presentations produced as part of the scheme in the paper, but
it is widely accepted within the computational group theory community that
Todd-Coxeter is the best choice for computing finite groups and monoids. Small
overlap monoids are infinite, and so techniques for small overlap monoids
cannot be applied here. A later version of this appendix will contain details
of the other approaches also.

The action of $M$ on $\rho$ is constructed as a specific type of digraph, which
is essentially a deterministic finite state automata without accept states. 
We refer to such digraphs as \textit{word graphs}. Suppose that $\Gamma$ is the
word graph of the action of $M$ on $\rho$. Then, roughly speaking, the
Todd-Coxeter algorithm produces a sequence of word graphs $\Gamma_0, \Gamma_1,
\ldots$ that approximate the word graph $\Gamma$. The initial word graph
$\Gamma_0$ has a single node $\varepsilon$ corresponding to the empty word, and
no edges. For every $i\geq 0$, there is a homomorphism from the graph
$\Gamma_{i}$ to $\Gamma_{i+1}$. The sequence $\Gamma_0, \Gamma_1, \ldots$ also
converges to $\Gamma$ in a certain precise categorical sense, even if $\Gamma$
is infinite. As such if a word $w\in A ^ *$ labels a path in $\Gamma_i$, then
$w$ labels a path in every $\Gamma_j$ where $j\geq i$. It follows that if $w
\in S$ labels a path starting and ending at node $\varepsilon$ in any
$\Gamma_i$, then $w$ also labels a path starting and ending at node
$\varepsilon$ in $\Gamma_j$ for every $j\geq i$. As such to determine whether
$w\in S$ encodes $0$, it suffices to find any approximation $\Gamma_i$ of
$\Gamma$ where $w$ labels a path starting and ending at $\varepsilon$.

\begin{figure}
  \centering
  \includegraphics[width=0.8\textwidth]{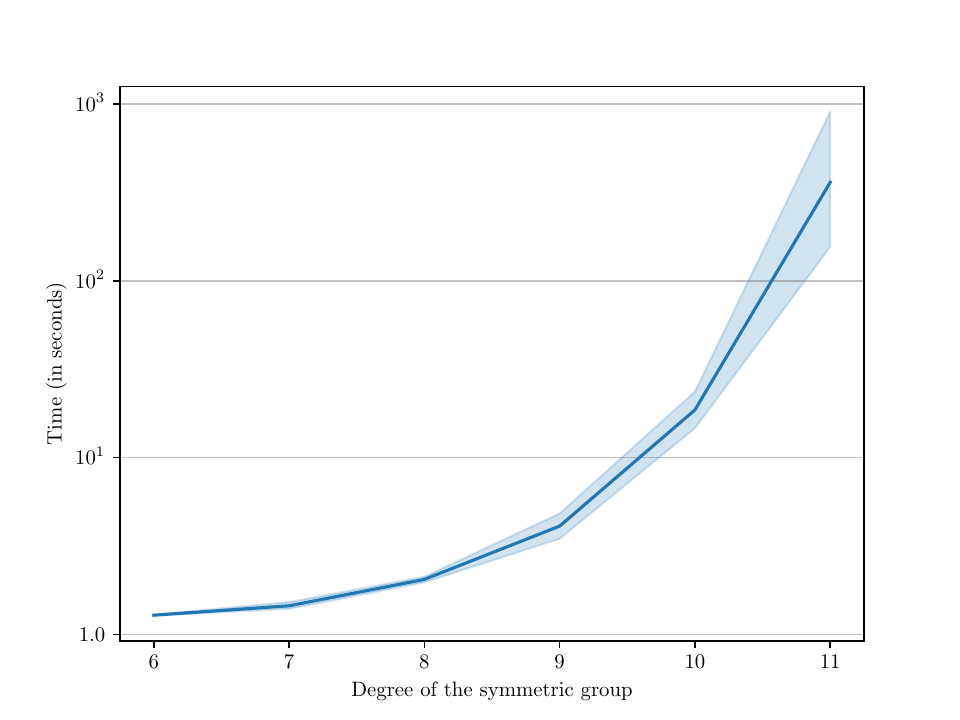}
  \caption{Time versus symmetric group degree for the non-semidirect product
  schemes }\label{figure-time-non-semidirect-product}
\end{figure}

We have considered first a series of challenges based on presentations of the
symmetric group $S_n$ for various values of $n$. These do not involve a
semidirect product construction, and are thus much less secure than the examples
obtained using the recommended parameters in \S\ref{sec-parameters} of the
paper. In Figure~\ref{figure-time-non-semidirect-product}, we have plotted the
time taken for the approach above to correctly decode at least $90\%$ of values
in the challenge words $S$ by comparison with the provided solutions $T$. The
$x$-axis corresponds to $n$.

All of the benchmarks were run on a server with a 2.20GHz Intel Xeon CPU with 16
cores, and 128GB of memory. 
Figure~\ref{figure-time-non-semidirect-product}
appears to show that both the amount of time and memory required to successfully
break these challenges increase exponentially. These figure represent an upper
bound on the run time for these computations. It is possible that by tweaking
the settings for the Todd-Coxeter algorithm in \libsemigroups that these times
could be decreased to some extent, but it seems unlikely that this would change
the overall exponential behaviour. Also we are solving a strictly easier
problem than an actual attacker would have to because we are provided with the
solution $T$. This allows us to accurately measure the progress of the
computation, and to terminate it early. Without the solution $T$, it is
significantly harder to measure how close to completion the algorithm is. 

There were several challenges that we could not break, using the technique
outlined above or in any other way, starting with the challenges created as above for $n=12$. Extrapolating the above curve could lead one to predict that the scheme could still be broken in a matter of hours, but in practice things appear much more complicated.

Besides, we have attempted to break further examples involving a
semidirect product $S_n \rtimes S_n$, as recommended in the
paper. We have not yet been able to break any of these as soon as $n>7$,
and certainly not for $n=11$.

To the best of our knowledge at the time of writing, \libsemigroups is the only
implementation of Todd-Coxeter that provides access to the approximation word
graphs $\Gamma_i$ in the sequence converging to $\Gamma$. It might still be
possible to use other implementations of Todd-Coxeter (such as those in \GAP or
\ACE) to break some of the challenges. However, there are also several
disadvantages to these tools: they were written in the 90s and 00s with the
computational resources of the time in mind; they were not designed to cope
with presentations as large as those produced by the scheme in the paper. 



\bibliography{info}
\bibliographystyle{amsplain}

\end{document}